\documentclass[12pt,preprint]{aastex}

\usepackage{natbib}
\usepackage{lineno}
\usepackage[figuresright]{rotating}
\usepackage{lscape}

\shorttitle{Evolving spectral-timing properties of a type-I X-ray burst of 4U~1730--22}
\shortauthors{Chen et al.}

\begin{document}


\title{
Return of 4U~1730--22 after 49 years silence: the peculiar burst properties  of the 2021/2022 outbursts observed by Insight-HXMT
}

\author{Yu-Peng Chen\textsuperscript{*}}
\email{chenyp@ihep.ac.cn}
\affil{Key Laboratory for Particle Astrophysics, Institute of High Energy Physics, Chinese Academy of Sciences, 19B Yuquan Road, Beijing 100049, China}

\author{Shu Zhang\textsuperscript{*}}
\email{szhang@ihep.ac.cn}
\affil{Key Laboratory for Particle Astrophysics, Institute of High Energy Physics, Chinese Academy of Sciences, 19B Yuquan Road, Beijing 100049, China}

\author{Long Ji\textsuperscript{*}}
\email{jilong@mail.sysu.edu.cn}
\affil{School of Physics and Astronomy, Sun Yat-Sen University, Zhuhai, 519082, China}

\author{Shuang-Nan Zhang}
\affil{Key Laboratory for Particle Astrophysics, Institute of High Energy Physics, Chinese Academy of Sciences, 19B Yuquan Road, Beijing 100049, China}
\affil{University of Chinese Academy of Sciences, Chinese Academy of Sciences, Beijing 100049, China}

\author{Peng-Ju Wang}
\affil{Key Laboratory for Particle Astrophysics, Institute of High Energy Physics, Chinese Academy of Sciences, 19B Yuquan Road, Beijing 100049, China}
\affil{University of Chinese Academy of Sciences, Chinese Academy of Sciences, Beijing 100049, China}

\author{Ling-Da Kong}
\affil{Key Laboratory for Particle Astrophysics, Institute of High Energy Physics, Chinese Academy of Sciences, 19B Yuquan Road, Beijing 100049, China}
\affil{ Institut f\"{u}r Astronomie und Astrophysik, Kepler Center for Astro and Particle Physics, Eberhard Karls Universit\"{a}t, Sand 1, D-72076 T\"{u}bingen, Germany}

\author{Zhi Chang}
\affil{Key Laboratory for Particle Astrophysics, Institute of High Energy Physics, Chinese Academy of Sciences, 19B Yuquan Road, Beijing 100049, China}

\author{Jing-Qiang Peng}
\affil{Key Laboratory for Particle Astrophysics, Institute of High Energy Physics, Chinese Academy of Sciences, 19B Yuquan Road, Beijing 100049, China}
\affil{University of Chinese Academy of Sciences, Chinese Academy of Sciences, Beijing 100049, China}

\author{Qing-Cang Shui}
\affil{Key Laboratory for Particle Astrophysics, Institute of High Energy Physics, Chinese Academy of Sciences, 19B Yuquan Road, Beijing 100049, China}
\affil{University of Chinese Academy of Sciences, Chinese Academy of Sciences, Beijing 100049, China}

\author{Jian Li}
\affil{CAS Key Laboratory for Research in Galaxies and Cosmology, Department of Astronomy, University of Science and Technology of China, Hefei 230026, China}
\affil{School of Astronomy and Space Science, University of Science and Technology of China, Hefei 230026, China}

\author{Zhao-Sheng Li}
\affil{Key Laboratory of Stars and Interstellar Medium, Xiangtan University, Xiangtan 411105, Hunan, P.R. China}

\author{Lian Tao}
\affil{Key Laboratory for Particle Astrophysics, Institute of High Energy Physics, Chinese Academy of Sciences, 19B Yuquan Road, Beijing 100049, China}

\author{Ming-Yu Ge}
\affil{Key Laboratory for Particle Astrophysics, Institute of High Energy Physics, Chinese Academy of Sciences, 19B Yuquan Road, Beijing 100049, China}

\author{Jin-Lu Qu}
\affil{Key Laboratory for Particle Astrophysics, Institute of High Energy Physics, Chinese Academy of Sciences, 19B Yuquan Road, Beijing 100049, China}
\affil{University of Chinese Academy of Sciences, Chinese Academy of Sciences, Beijing 100049, China}

\begin{abstract}
After in quiescence for 49 years, 4U~1730--22 became active and had two outbursts in 2021 \& 2022; ten thermonuclear X-ray bursts were detected  with Insight-HXMT.
Among them, the faintest burst showed a double-peaked profile,  placing the source as the 5th accreting neutron star (NS)   exhibiting double/triple-peaked type-I X-ray bursts; the other bursts showed photospheric radius expansion (PRE).
The  properties of double-peaked non-PRE burst indicate that it could be related to a stalled burning front.
For the five bright PRE bursts, apart from the emission from the neutron star (NS) surface,  we find the residuals both in the soft ($<$3 keV) and hard ($>$10 keV) X-ray band.
Time-resolved spectroscopy reveals that the excess can be attributed to an enhanced pre-burst/persistent emission or the Comptonization of the burst emission by the corona/boundary-layer.
We  find,  the burst emission shows a rise until the photosphere touches down to the NS surface rather than the theoretical predicted  constant Eddington luminosity.
The shortage of  the burst emission in the early rising phase is beyond  the occlusion by the disk.
We speculate that
the findings above correspond to that the obscured part (not only the lower part) of the NS surface is exposed to the line of sight due to the  evaporation of the obscured material by the burst emission, or the  burst emission is anisotropic ($\xi>1$) in the burst early phase.
In addition, based on the average  flux of  PRE
bursts at their touch-down time,
we derive a distance estimation as 10.4 kpc.

\end{abstract}

\section{Introduction}

Until June of 2022, there were three outbursts observed from 4U~1730--22: lasting $\sim$ 6 months  in 1972 detected by Uhuru \citep{Cominsky1978,Forman1978}, lasting 1 month  in 2021 and lasting 3 months  in 2022,  and all three with peak flux of 100 mCrab in soft X-ray band.
In the interval between the first two outbursts, a bright  X-ray emission in its quiescent state was observed by Chandra, which indicates its  neutron star's (NS) nature \citep{Tomsick2007} and  its distance estimation of $\sim$ 10 kpc.
In the 2nd outburst, 49 years after the 1st one, thermonuclear bursts were detected in 4U~1730--22 by NICER, which  identified as   NS X-ray binary \citep{Li2022}.
Its spin is around  $\nu$=584.65 Hz \citep{Li2022}, based on the burst oscillation detection by NICER.

Type-I X-ray bursts, also named thermonuclear X-ray bursts, are triggered by  unstable thermonuclear burning of the  accumulated accretion fuel from a  low-mass X-ray binary (LMXB)  hosting a NS (for reviews, see \citealp{Lewin,Cumming,Strohmayer,Galloway}).
Bursting behavior is known to be extremely variable and violent,  and most bursts manifest as a fast-rise (seconds), an exponential decay ($\sim$ 10 s to  minutes) and a peak luminosity up to the  Eddington luminosity.
Most bursts are single-peaked, except the brightest ones which show  photospheric radius expansion (PRE; due to radiation pressure).

For the PRE bursts, the radiation pressure of the burning exceeds the NS gravitational force in the photosphere, resulting in an increase of the photosphere radius and a decrease of the photosphere temperature in an  adiabatic expansion;
as the   expansion ceases,  the lifted photosphere drops on the NS surface and   an increase of the
photosphere temperature  is shown in an  adiabatic compression \citep{Grindlay1980}.

The above spectral shift causes a dip in the temperature, a spike in the radius and a plateau  in the luminosity in the PRE phase.
In the lightcurves,
a single-peaked structure is typically observed with   soft X-ray instruments, e.g., RXTE/PCA, Swift/XRT, Chandra, XMM-Newton,  NICER and Insight-HXMT/LE, but a double-peaked structure is often seen with  hard X-ray instruments because of the passband limitation, e.g., INTEGRAL, Swift/BAT, RXTE/HEXTE and Insight-HXMT/HE.

The vast majority of the bursts with luminosity below the Eddington limit show a single peak in the lightcurves.
Non-PRE bursts with double-peaked or triple-peaked structures have been detected in four bursters, e.g., 4U~1636--536 \citep{Bhattacharyya2006,Zhang2009}, 4U~1608--52 \citep{Penninx1989,Guver2021}, GX~17+2 \citep{Kuulkers2002}, and GRS~1741.9--2853 \citep{Pike2021}.
Potential explanations include multiple generations/release of thermonuclear energy,  absorption/scattering from an accretion-disk corona, and flame spread stalling on the NS surface \citep[e.g., ignites at high latitude but stalls on the equator;][]{Pike2021}.

Since the bursts occur on the NS surface, the interplay between the NS surface emission and the accretion environment should be taken into account. In recent ten years, among thousands of observed bursts from the 118  bursters\footnote{https://personal.sron.nl/$\sim$jeanz/bursterlist.html},  impacts observed on the accretion process by bursts have been observed, i.e., an enhancement/deficit \citep{Worpel2013, Worpel2015,Bult2021,Ji2014} at soft X-ray band, a shortage at hard X-ray band \citep{maccarone2003, Chen2012, Ji2013},   a bump peaking at 20--40 keV and/or discrete emission by reflection from accretion disk \citep{int2013, Ball2004, Keek2014}.

In this work, using a broad energy band capabilities of Insight-HXMT in 1--50 keV, we study ten bursts from 4U~1730--22: one double-peaked burst and nine PRE bursts.
The present paper focuses on the nature of these bursts and also examines the effect of the burst emission on the accretion environment using a variable persistent flux method and Comptonization of the burst emission by the surrounding hot electrons. We describe the observations and data reduction  in Section 2,  present our analysis methods, spectral results on outburst and  spectral/temporal properties of the bursts in Section 3. Finally,  a discussion and understanding of the above results are given in Section 4.

.

\section{Observations and Data Reduction}

\subsection{Insight-HXMT}
Hard X-ray Modulation Telescope (HXMT, also dubbed as Insight-HXMT, \citealp{Zhang2020}) excels in its broad energy band (1--250 keV) and a large effective area in the hard X-rays energy band.
It carries three collimated telescopes: the High Energy X-ray Telescope (HE; poshwich NaI/CsI, 20–250 keV, $\sim$ 5000 cm$^2$), the Medium Energy X-ray Telescope (ME; Si pin detector, 5–40 keV, 952 cm$^2$) and the Low Energy X-ray telescope (LE; SCD detector, 1–12 keV, 384 cm$^2$).
Under the quick read-out system of Insight-HXMT detectors, there is little pile-up effect at the burst peak.
Insight-HXMT Data Analysis software (HXMTDAS) v2.05\footnote{http://hxmtweb.ihep.ac.cn/} is used to analyze the data.

\begin{figure}[t]
\centering
\includegraphics[angle=0, scale=0.8]{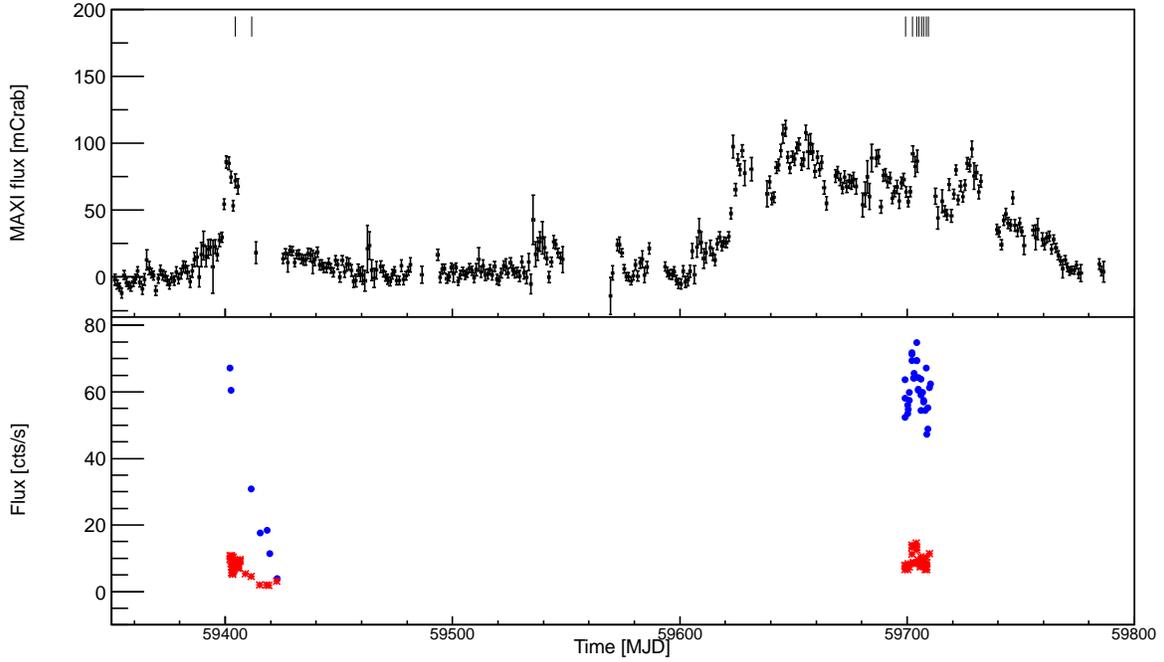}
  \caption{Top panel: daily light curves of 4U~1730--22 by MAXI (black)  during the outbursts in 2021 and 2022 in 2–20 keV. The  bursts are indicated by vertical lines. Bottom panel: light curves of 4U~1730--22 by LE (blue) and ME (red), which are rebinned by one obsid ($\sim$ 10000 s).
  }
\label{fig_outburst_lc}
\end{figure}

As shown in Figure \ref{fig_outburst_lc},  for the two outbursts in 2021 and 2022,
Insight-HXMT have observed 4U~1730--22 with 74 observations ranging from P041401100101-20210707-01-01 to P051400201402-20220513-02-01  with a total   observation time of 184 ks.
These observations covered the peak/decay phase of the outburst in 2021  and the plateau of the  outburst in 2022.


We note that the default good-time-interval (GTI) selection criteria of LE are very conservative because of the influence of light leaks.
To obtain a complete sample of bursts, lightcurves are extracted without filtering GTIs.
Burst-like fluctuations that may be caused by a sharp variation of the background, when the telescope passes the South Atlantic Anomaly (SAA), are excluded.

As shown in Table \ref{tb_burst}, 10 bursts are found in ME and HE data with a peak flux $\sim$300--600 cts/s and $\sim$100--200 cts/s;  among them, 6 bursts are also found in LE data with a peak flux $\sim$300--1200 cts/s.

For each burst, we use the time of the ME flux peak as a reference (0 s in Figure \ref{fig_burst_lc}) to produce lightcurves and spectra.
We extract time-resolved spectra of LE, ME and HE with a bin size of 0.5 s starting from the onset of each burst.
As a conventional procedure, the pre-burst emission (including the persistent emission and the instrumental background) is extracted, which is taken as the background when fitting the  spectra during bursts.
  In practice, for each burst, we define the time interval between 70 and 20 seconds before the burst peak as the time window of the pre-burst emission, i.e., [-70 s, -20 s].

For the outburst in 2022, the last eight bursts locates at the plateau of the outburst.
In this period, the overlapped observations between Insight-HXMT and NICER are P051400200102-20220430-01-01 and 4639010134, respectively. Thus, we get 600 s and 2300 s GTI of LE and ME.
The HE spectrum is not involved in the joint spectral fitting of the persistent emission, since the HE detection  falls below the systematic error of the background model.
Please notice that the HE spectra are involved in the bursts analysis, since  the peak fluxes of the burst detected by HE are much brighter than the persistent emission.


The other results, e.g, the persistent spectra, background and net lightcurves are obtained  following the recommended procedure of the Insight-HXMT Data Reduction, which are screened with the standard criterion included in Insight-HXMT pipelines: lepipeline, mepipeline and hepipeline. 

For the persistent emission spectral fitting of LE and ME, the energy bands are chosen to be 2--7 keV and 8--20 keV.
The spectra are rebinned by ftool ftgrouppha \citep{Kaastra2016} optimal binning algorithm with a minimum of 25 counts per grouped bin.

The LE background model works only in a certain temperature range \citep{Li2020}. This leads to some uncertainties below 2 keV caused by the electronic noise when the temperature exceeds this range after the middle year of 2019. During a burst with a time-scale of tens of seconds, the temperature fluctuation of LE is so small that can be neglected.
The resulting electronic noise of the pre-burst spectrum is the same as that of burst spectra.
Therefore, the influence of the electronic noise can be canceled out when we take the pre-burst spectrum as the background of burst spectra.
In this case, the energy band of LE can be extended to 1--10 keV in the burst analysis.

The ME and HE energy bands used in burst spectral fitting are   8--30 keV and 25--50 keV, respectively.
The slices of burst spectra of LE, ME and HE are rebinned by ftool grppha with a minimum of 10 counts per grouped bin, based on the limited photons of the burst slice spectra due to the short exposure time. We added a systematic uncertainty of 1\% to the Insight-HXMT spectra to account for the systematic uncertainties in the detector calibrations \citep{Li2020}.

\subsection{NICER}

For the two outbursts,  NICER also performed high cadence  observations on  4U~1730--22.  
There are two bursts during the outburst in 2021, but there is no LE data for both of them.
Without the LE data, the canonical blackbody model could fit the burst spectra well, and  there is no need for adding another component during the fitting, e.g., the variable persistent emission.
Under this condition, only the persistent spectra of the outburst in 2022 are extracted for fitting,  and the derived parameters of the model is used to fit the burst spectra.

There are several NICER observations in the plateau of the outburst of 2022, and we choose an overlapped obsid of 4639010134 (Table \ref{tb_nicer_obsid}) to   joint  fit the spectra of  Insight-HXMT and NICER.
The observation has a GTI $\sim$ 450 s and a count rate $\sim$ 600  cts/s in the 0.3–12 keV band.   

The NICER data are reduced using the pipeline tool nicerl2\footnote{https://heasarc.gsfc.nasa.gov/docs/nicer/nicer\_analysis.html} in NICERDAS v7a with the standard NICER filtering and using ftool XSELECT to extract lightcurves and spectra.
The background is estimated using the tool nibackgen3C50 \citep{Remillard2022}. The Focal Plane Module (FPM) No. 14 and 34 are removed from the analysis because of increased detector noise.
The response matrix files (RMFs) and ancillary response files (ARFs) are generated with the ftool nicerrmf and nicerarf.
The spectra are rebinned by ftool ftgrouppha \citep{Kaastra2016} optimal binning algorithm  with a minimum of 25 counts per grouped bin.

The tbabs model with Wilm abundances accounts for the ISM absorption in the spectral model \citep{Wilms2000}.
 The resulting spectra are analyzed using XSPEC \citep{Arnaud1996} version 12.11.1.
We added a systematic uncertainty of 1\% to the NICER spectrum.


\section{Analysis and Results}
\subsection{Fitting the joint Insight-HXMT/NICER spectrum of persistent emission}

We fit the joint NICER and Insight-HXMT (LE and ME) spectra with an absorbed convolution thermal Comptonization model (with   input photons contributed by the spectral component  diskbb), available as thcomp (a more accurate version of nthcomp) \citep{Zdziarski2020} in XSPEC, which is described by the optical depth $\tau$, electron temperature $kT_{\rm e}$, scattered/covering fraction $f_{\rm sc}$.
The  hydrogen column (tbabs in XSPEC)
accounts for both the line-of-sight column density and any intrinsic absorption near the source.
The seed photons are in the shape of diskbb since the thcomp model is a convolution model, and a fraction of Comptonization photons  are also given in the model.
Normalization constants are included during fittings to take into account the inter-calibrations of the instruments.
We keep the normalization factor of the NICER data with respect to the LE and ME  data to unity.

\begin{figure}[t]
\centering
      \includegraphics[angle=270, scale=0.3]{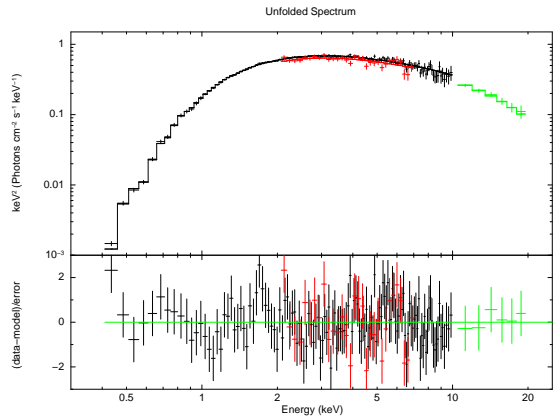}
 \caption{The spectral fit results of the  persistent emission by  NICER (black), LE (red) and ME (green)   with model cons*tbabs*thcomp*diskbb.
 }
\label{fig_outburst_spec}
\end{figure}
\begin{figure}[t]
\centering
\includegraphics[angle=0, scale=0.2]{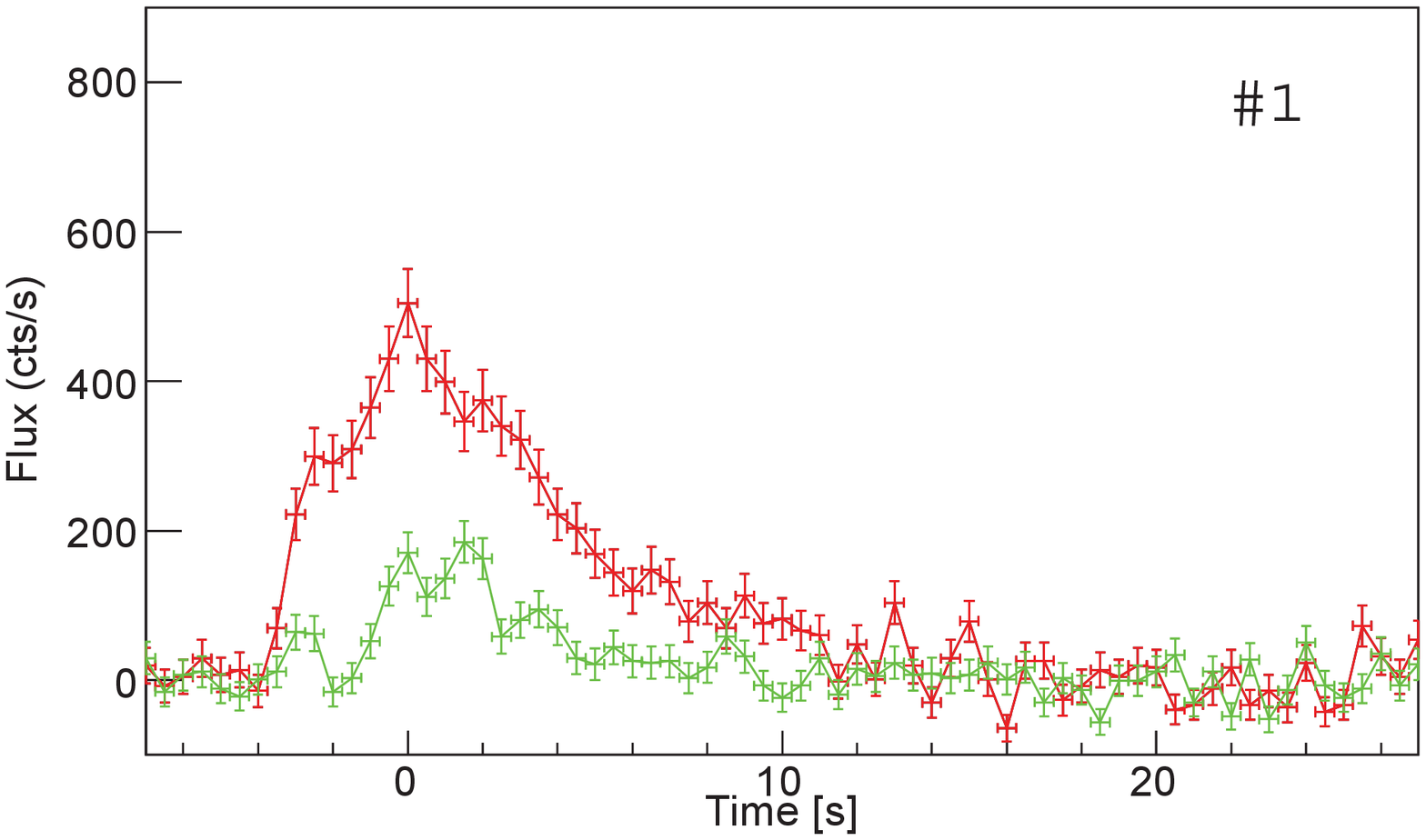}
\includegraphics[angle=0, scale=0.2]{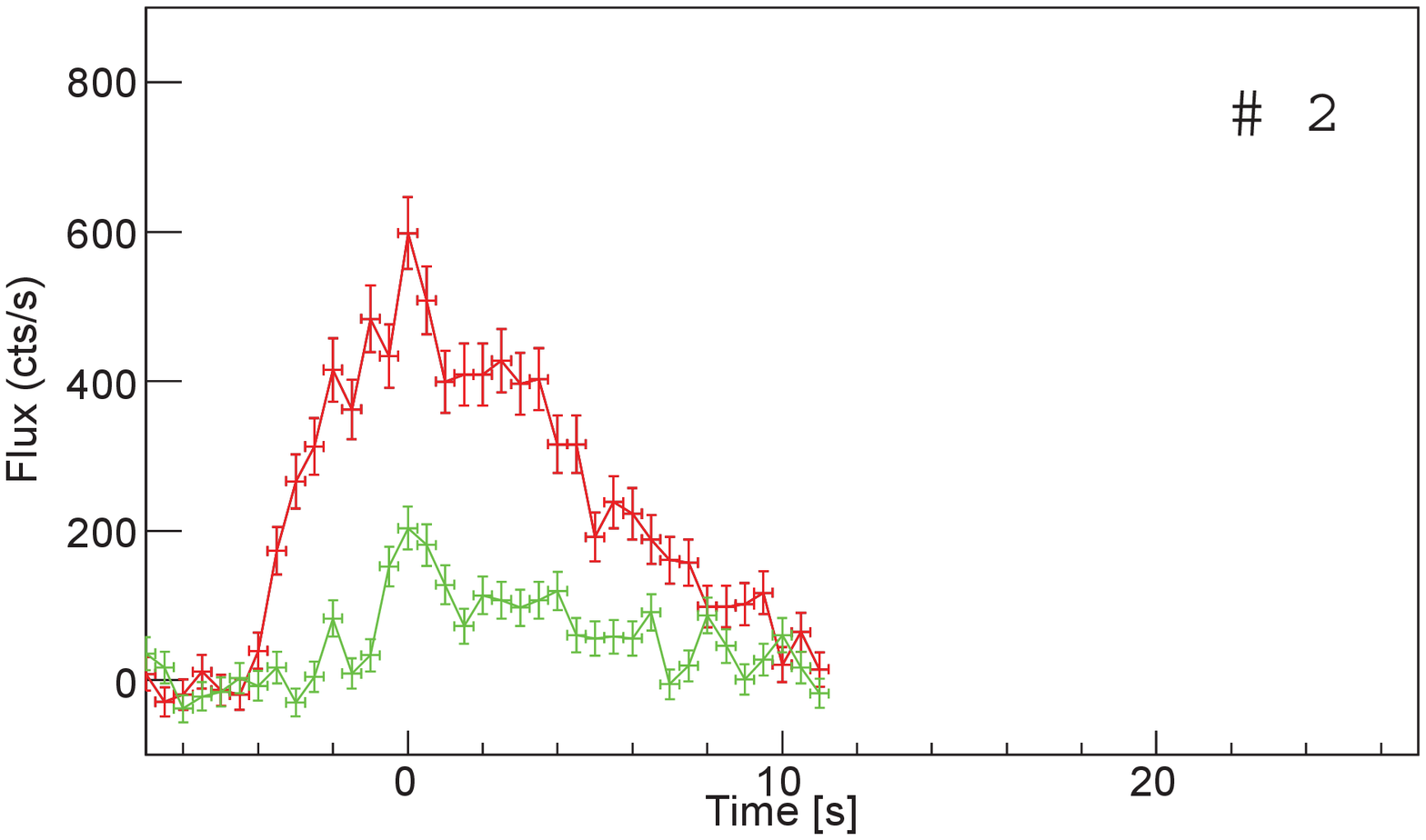}
\includegraphics[angle=0, scale=0.2]{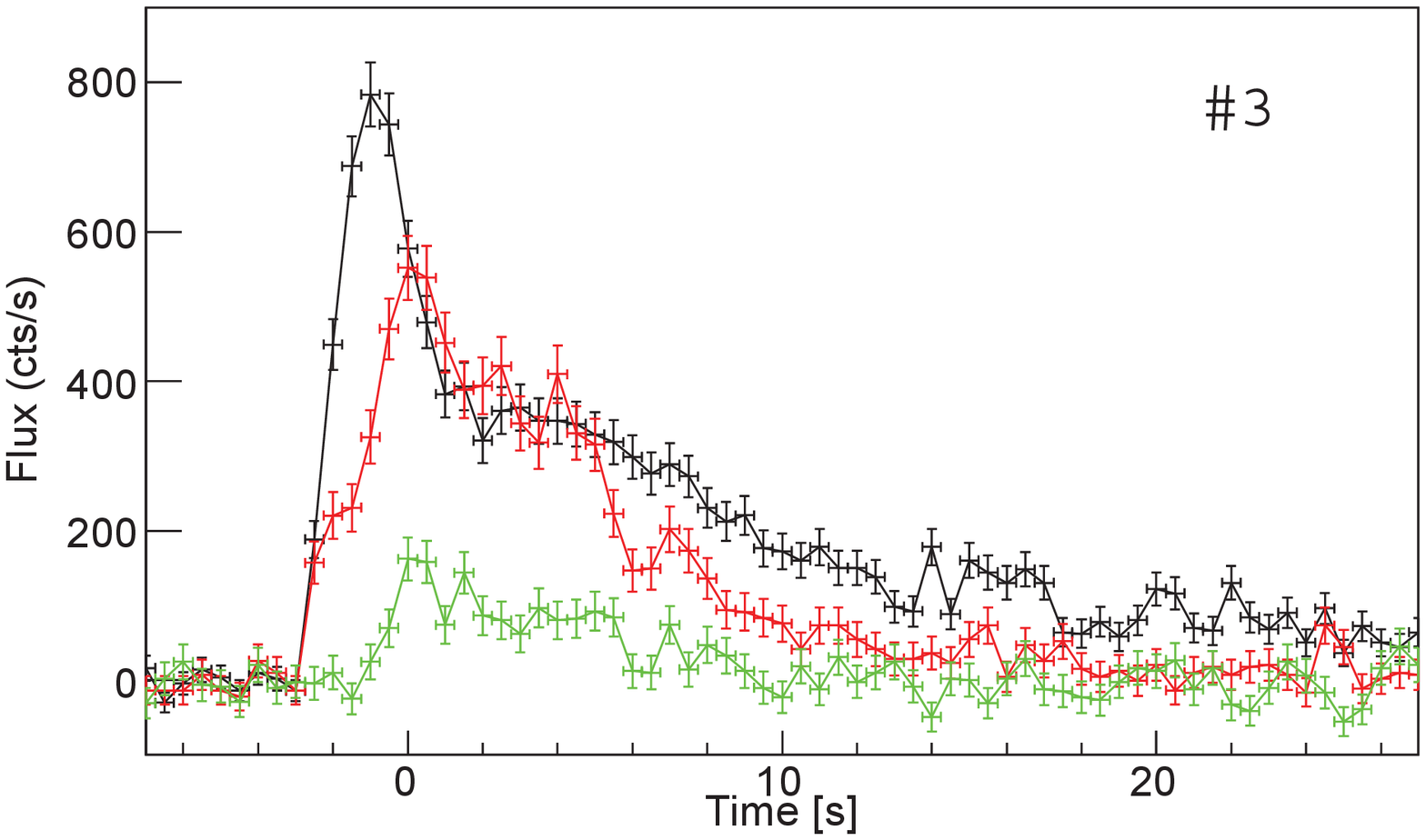}
\includegraphics[angle=0, scale=0.2]{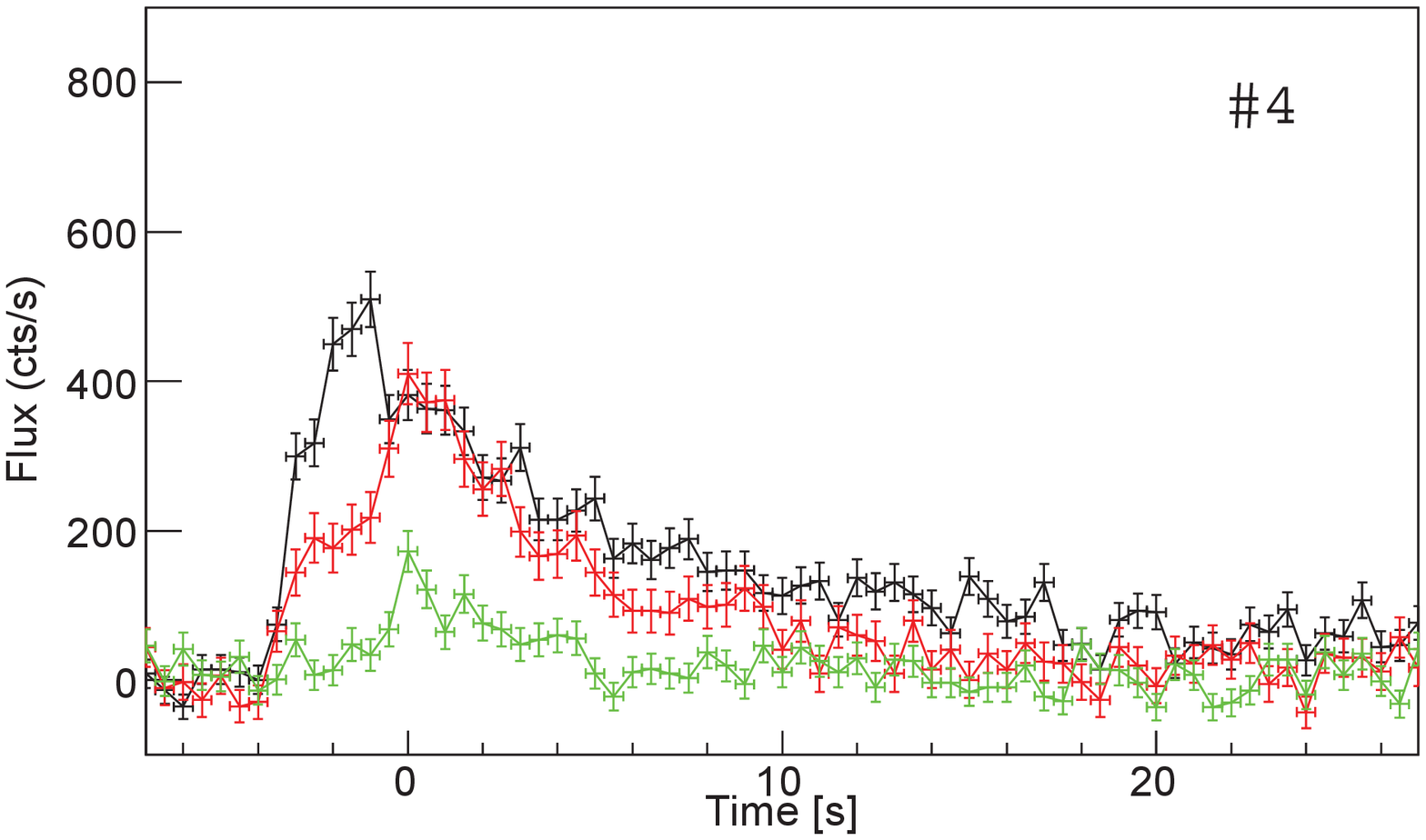}
\includegraphics[angle=0, scale=0.2]{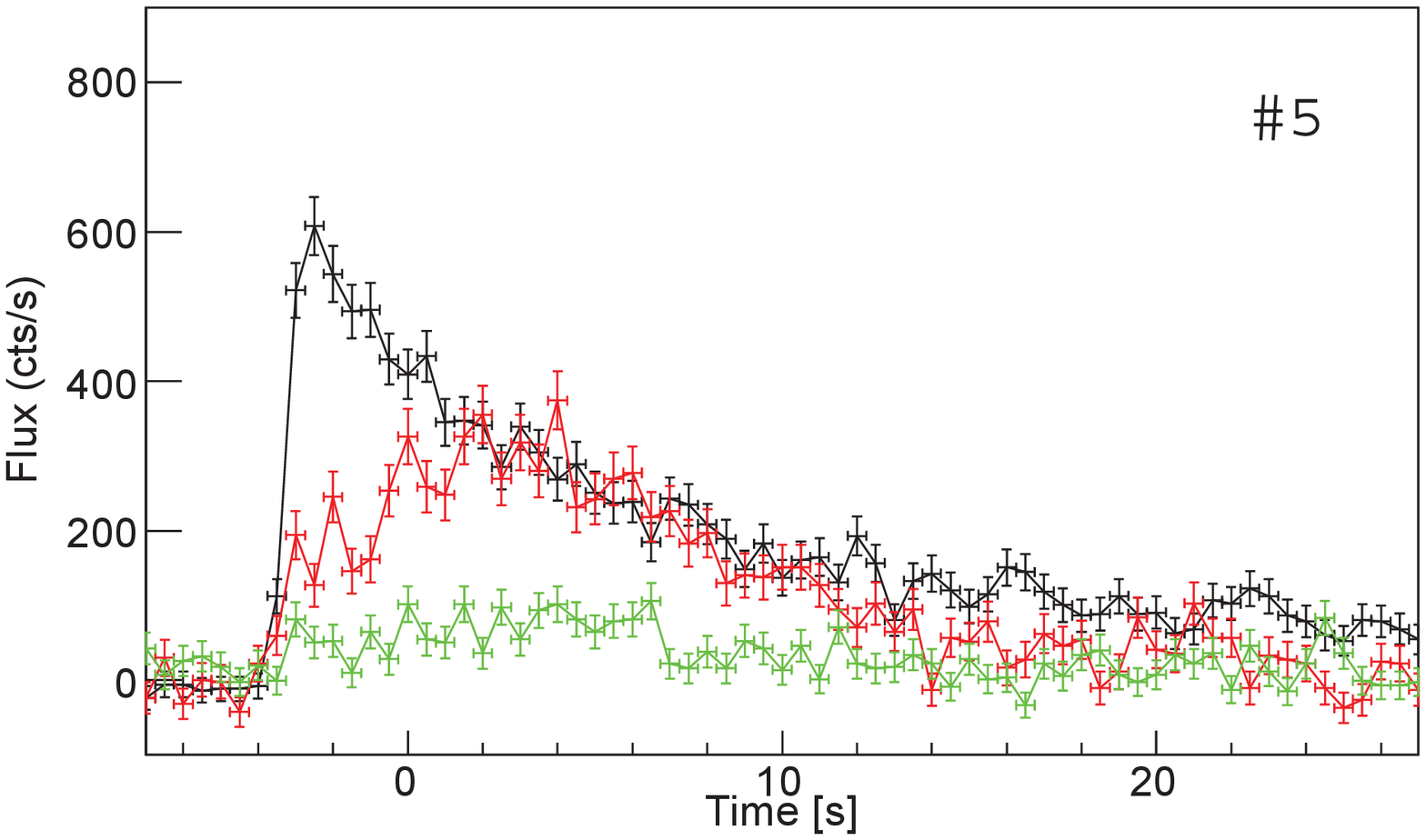}
\includegraphics[angle=0, scale=0.2]{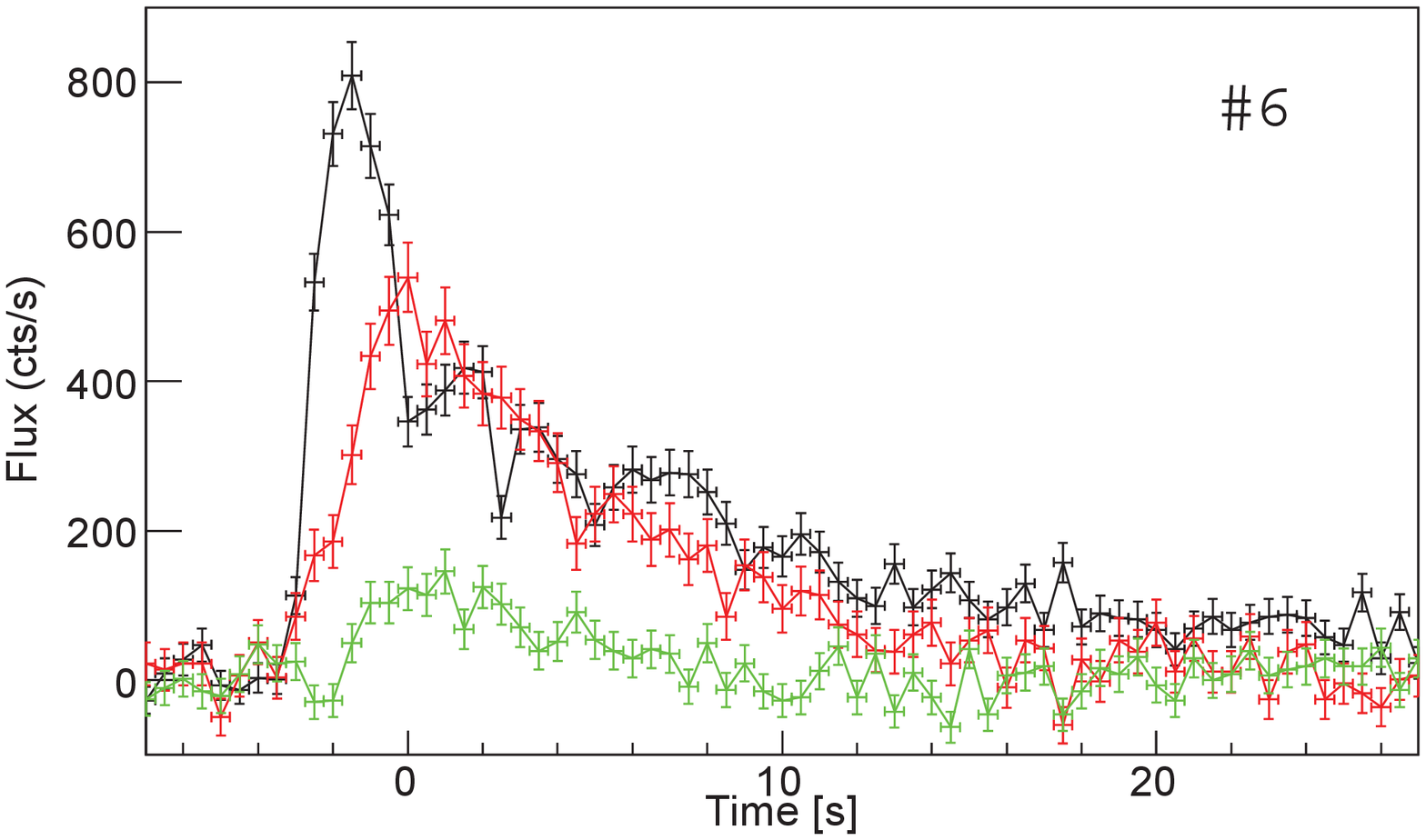}
\includegraphics[angle=0, scale=0.2]{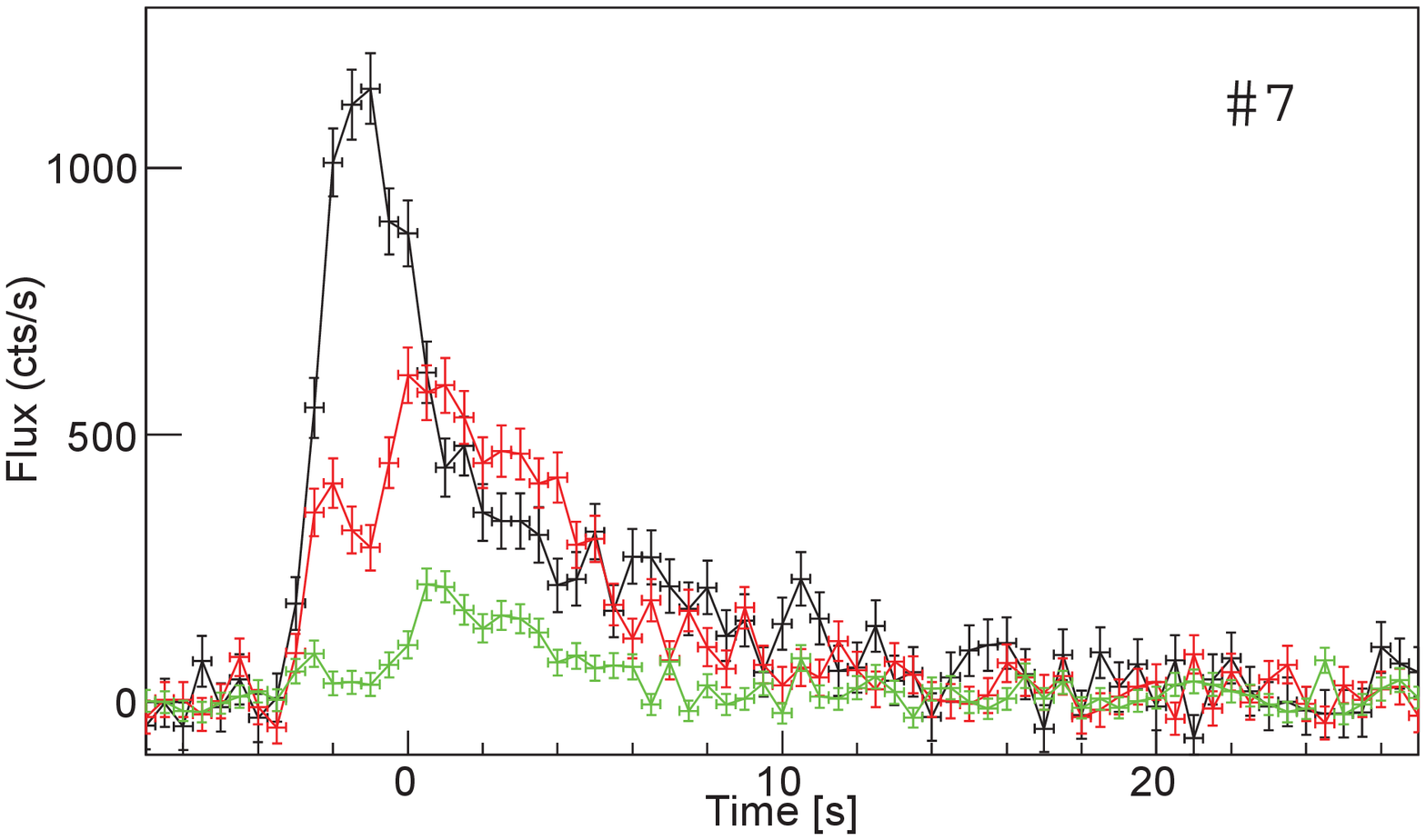}
\includegraphics[angle=0, scale=0.2]{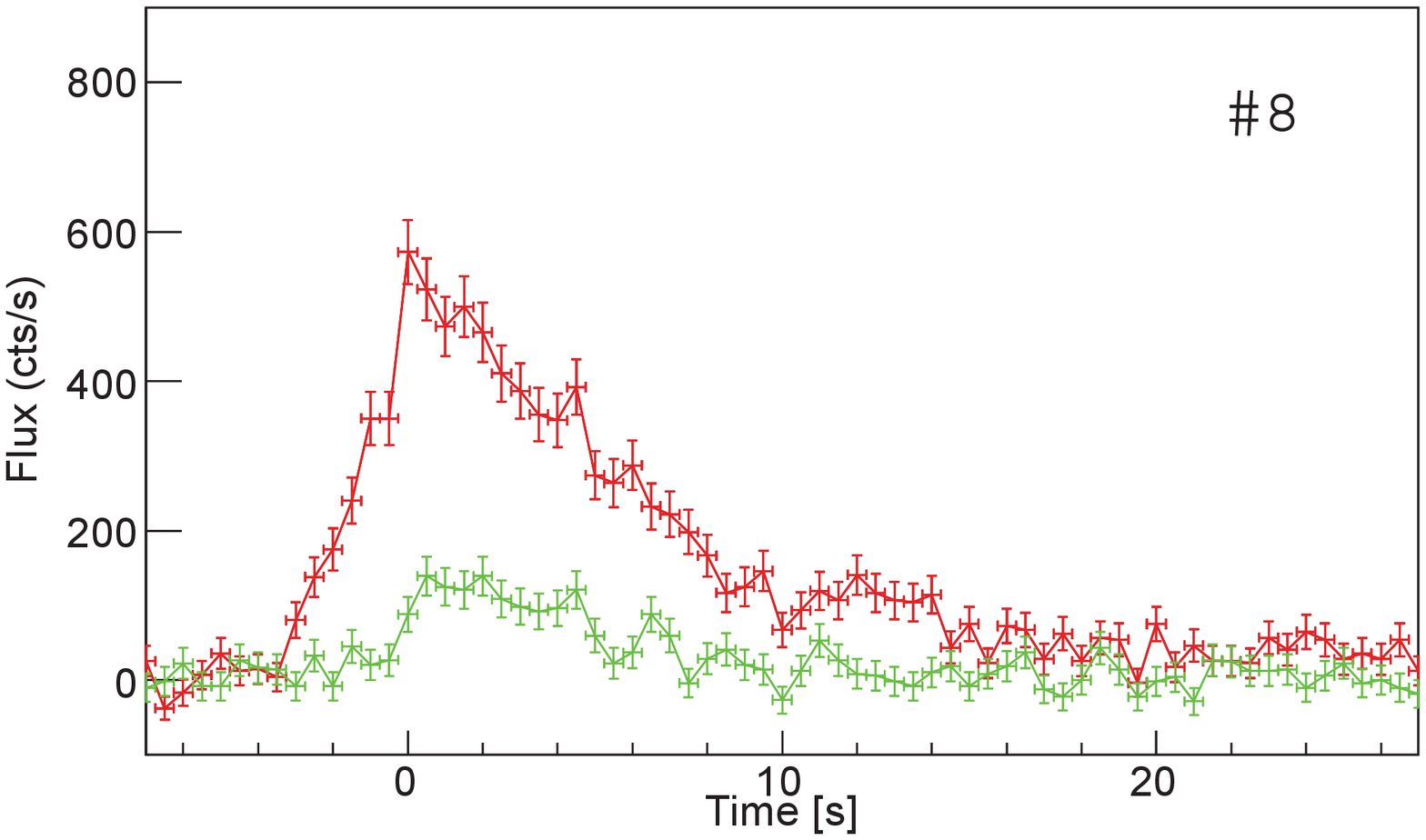}
\includegraphics[angle=0, scale=0.2]{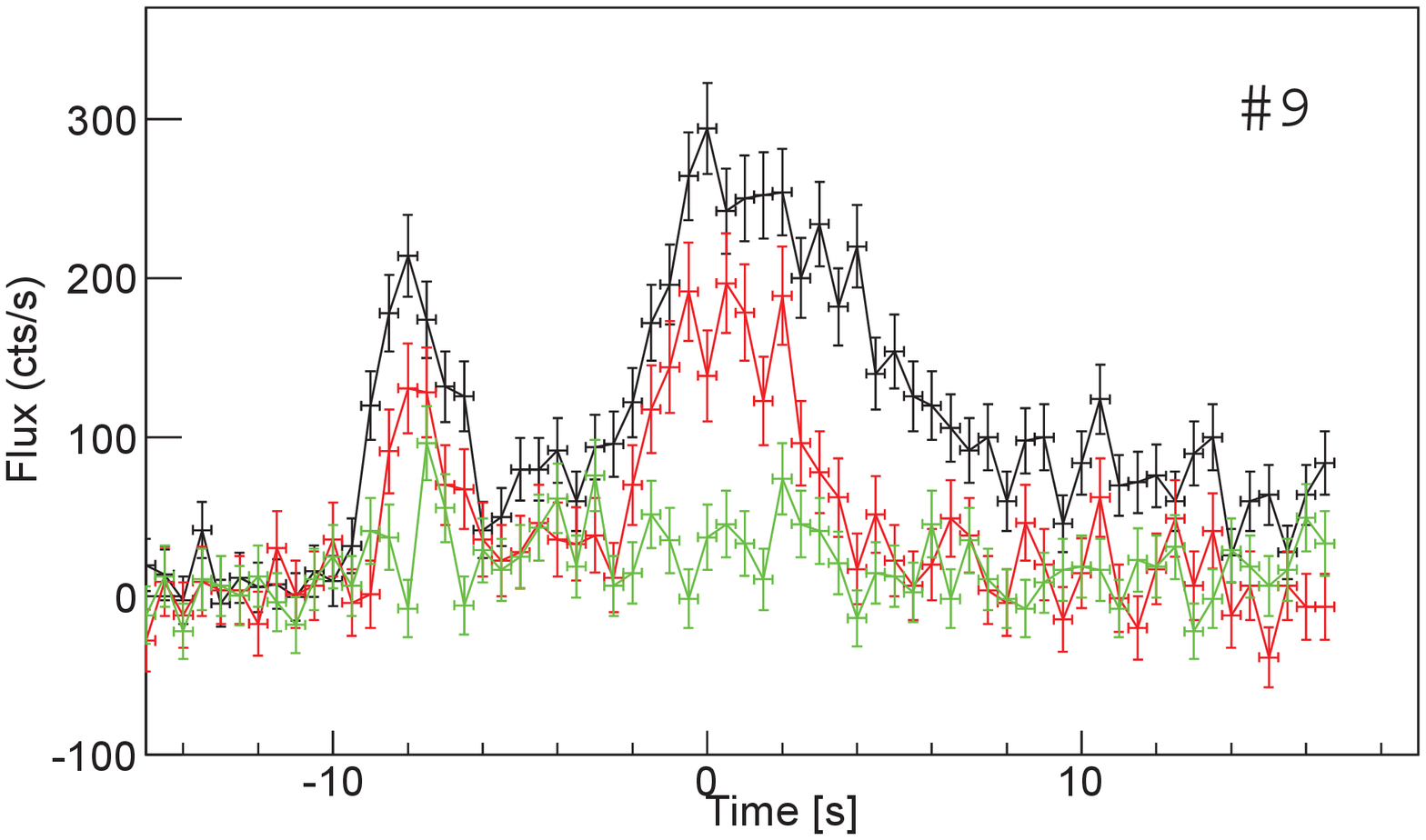}
\includegraphics[angle=0, scale=0.2]{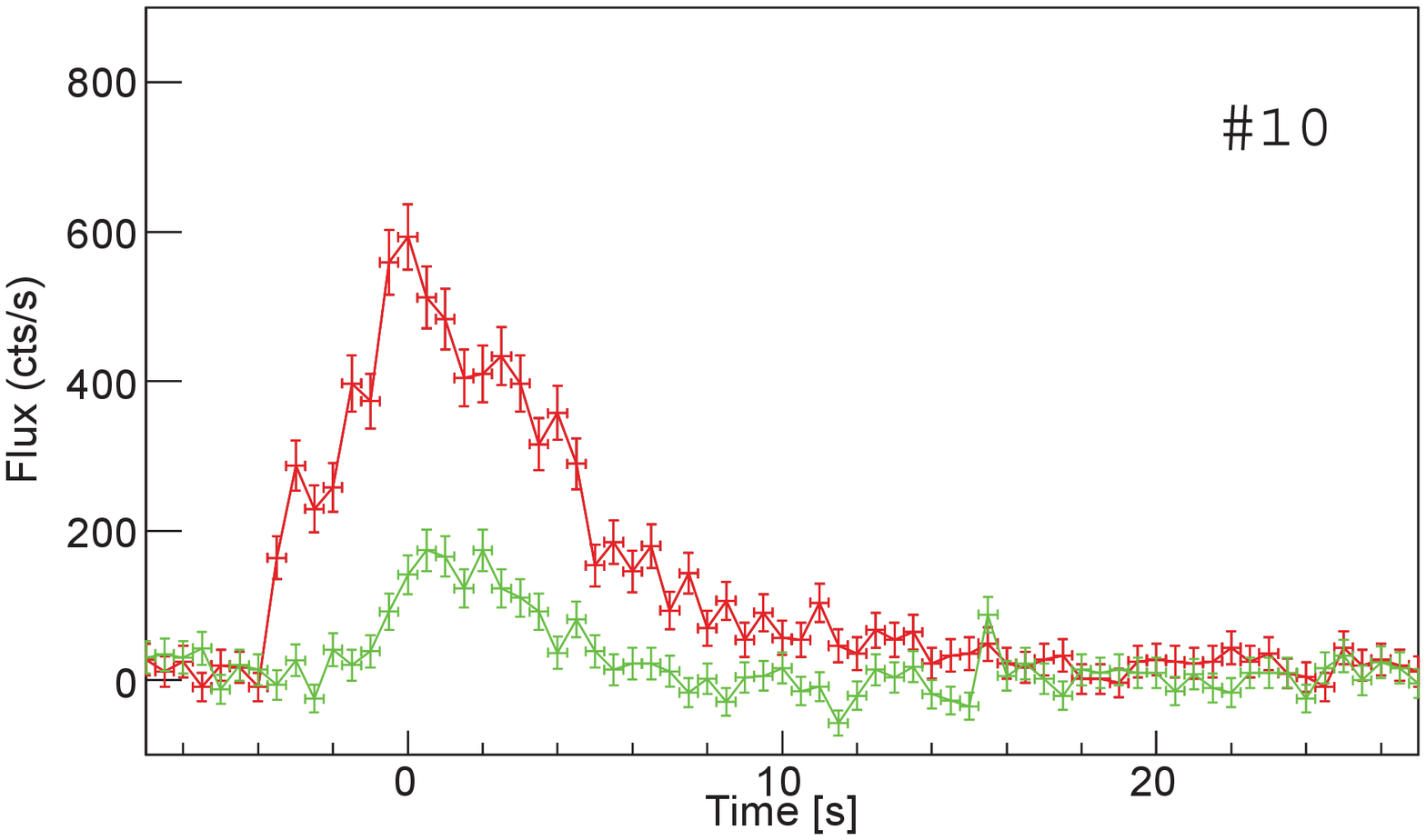}
 \caption{
 Lightcurves  with pre-burst emission subtracted of the 10 type-I X-ray bursts detected in the Insight-HXMT observation of 4U~1730--22 with time bin 0.5 s by LE (black), ME (red) and HE (green). The lightcurves of LE and ME are in  their full energy bands; the HE lightcurves result in 20--50 keV.
  }
\label{fig_burst_lc}
\end{figure}

Using the model above, we find an acceptable fit: reduced $\chi_{\upsilon}$=0.91 (d.o.f. 160; Figure \ref{fig_outburst_spec} and Table \ref{tb_persist_fit}), with the inner disc radius $R_{\rm diskbb}$ and  temperature $kT_{\rm in}$   found to be $\sim 19.3_{-1.6}^{+1.9}$ km (with a distance of 10 kpc and an inclination angle of 0 degree) and $0.68_{-0.34}^{+0.39}$ keV respectively.
Please note that the distance is given in the following pages based on the PRE bursts.
The thcomp parameters of the electron temperature $kT_{\rm e}$ and optical depth $\tau$ are $3.61^{+0.42}_{-0.57}$ keV and $7.8^{+1.0}_{-0.9}$.
 The scattered/covering fraction $f_{\rm sc}$ is derived as $0.929$.
However, when we extract  the  confidence region for $f_{\rm sc}$, the parameter is pegged at hard limit--1.
It is then frozen at 1; as expected, the results are consistent with each other within parameter's error bar.
The derived hydrogen column density $N_{\rm H}$ is $0.53\pm0.01\times 10^{22}~{\rm cm}^{-2}$.
The constants of LE and ME are 0.93$\pm0.01$ and 0.83$\pm$0.05, respectively.
The inferred bolometric flux in 0.01--1000 keV is
$3.08_{-0.03}^{+0.04}\times10^{-9}~{\rm erg}~{\rm cm}^{-2}~{\rm s}^{-1}$
corresponding to 20.5\% $L_{\rm Edd}$ at distance of 10 kpc, with $L_{\rm Edd}=1.8\times10^{38}$ erg/s.

The other scenario, i.e., substituting the diskbb component with a blackbody component in the aforementioned convolution model, is also attempted.
Taking this approach, spectral fits yield roughly the same thcomp parameters and reduced $\chi_{\upsilon}$=0.93  (the same d.o.f.).
However, the derived blackbody radius is $80\pm3$ km, which is far greater than the NS radius.

\subsection{Burst lightcurves by Insight-HXMT}

\subsubsection{The lightcurve of the double-peaked burst}

We show the LE/ME/HE lightcurves in Fig. \ref{fig_burst_lc} with a time resolution of 0.5 s.
The burst profiles  exhibit a  typical fast rise and slow (exponential) decay in the X-ray band.
For the faintest burst, burst \#9, there is a double-peaked structure with an interval of $\sim$ 10 s. Both for the lightcurves of  LE and ME, the peak flux of the first sub-burst is $\sim$2/3 of that of the second one. For HE, there is an enhancement in the lightcurve, and  no dip between the two sub-bursts.
The rising rate of the two sub-bursts is similar for the ME: 50 cts/s increases  every 0.5 s, but the decay of the first sub-burst is much faster than the second one.

\subsubsection{The lightcurve of the PRE burst}

For the five bursts with detection of LE, bursts \#3--\#7, the  lightcurves of  LE  present a single-peaked structure. Moreover, the hard X-rays (ME and HE)   lag  behind the soft X-rays (LE) by $\sim$ 1 s; the peak times of the ME and HE lightcurves are consistent with each other.
The brightest burst, burst \#7,  shows a double-peaked profile in ME and HE lightcurves,  which is a typical characteristic of a PRE burst.
For other bright bursts, e.g., bursts \#1, \#2, \#4, \#6, there are only  hints of another spike on the onset of the burst.

\subsection{Broad-band spectra of burst emission by Insight-HXMT}

When we fit the burst spectra, we estimate the background using the emission before the burst, i.e., assuming the persistent emission is unchanged during the burst. 
To account for the effective area calibration deviation, a constant is added to the model.
At the first attempt, for ME, the constant is fixed to 1, the others are variable during spectral fitting. The fits indicate that most of the constants of HE  and some of the constants of
LE are not convergent, owing to the low-significance data. Under this situation, the constants of LE \& HE are fixed at 1 for the combined-spectral fitting.

\subsubsection{Fit the spectra of bursts by the blackbody model }

\begin{figure}[t]
\centering
\includegraphics[angle=0, scale=0.2]{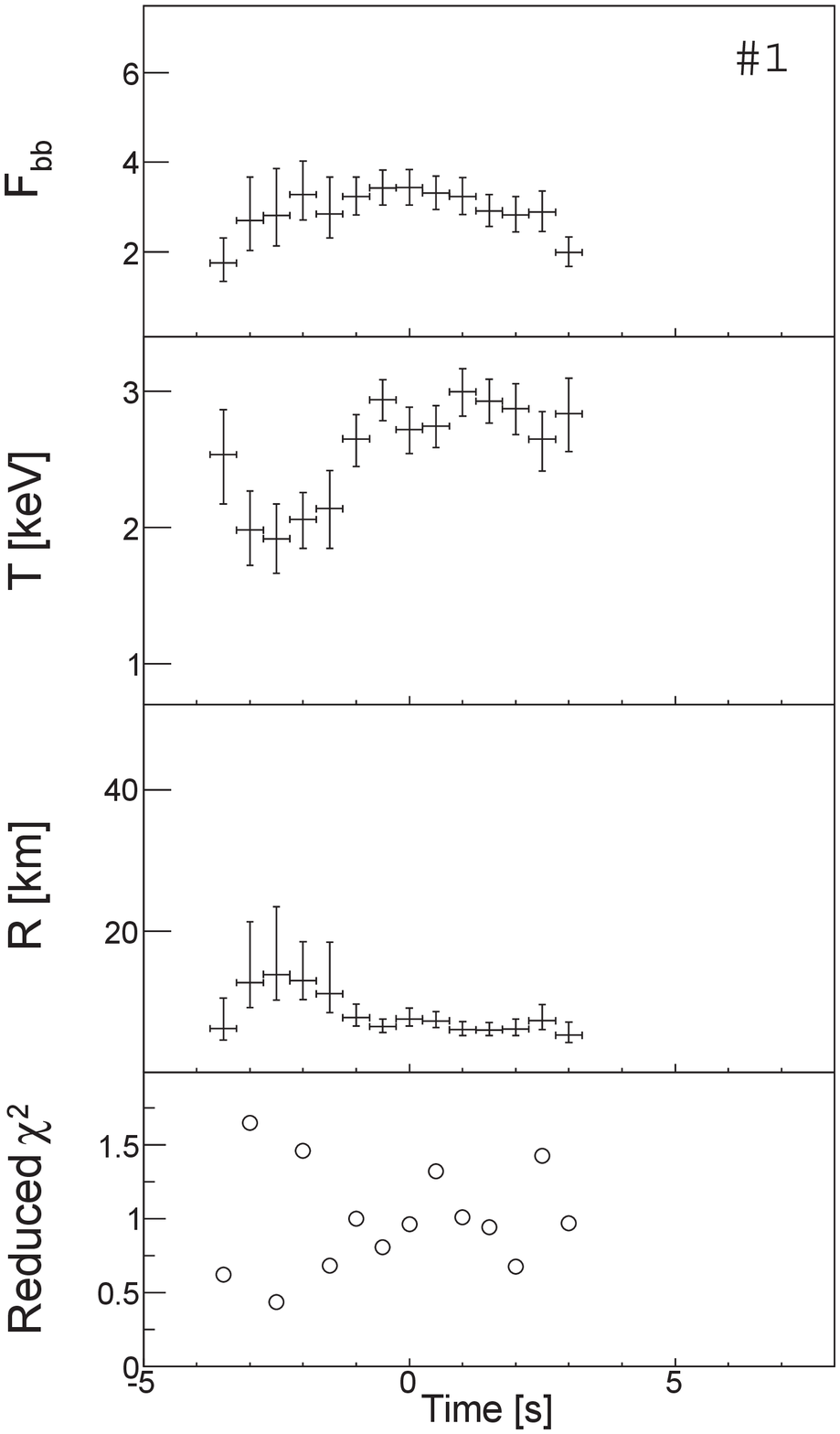}
\includegraphics[angle=0, scale=0.2]{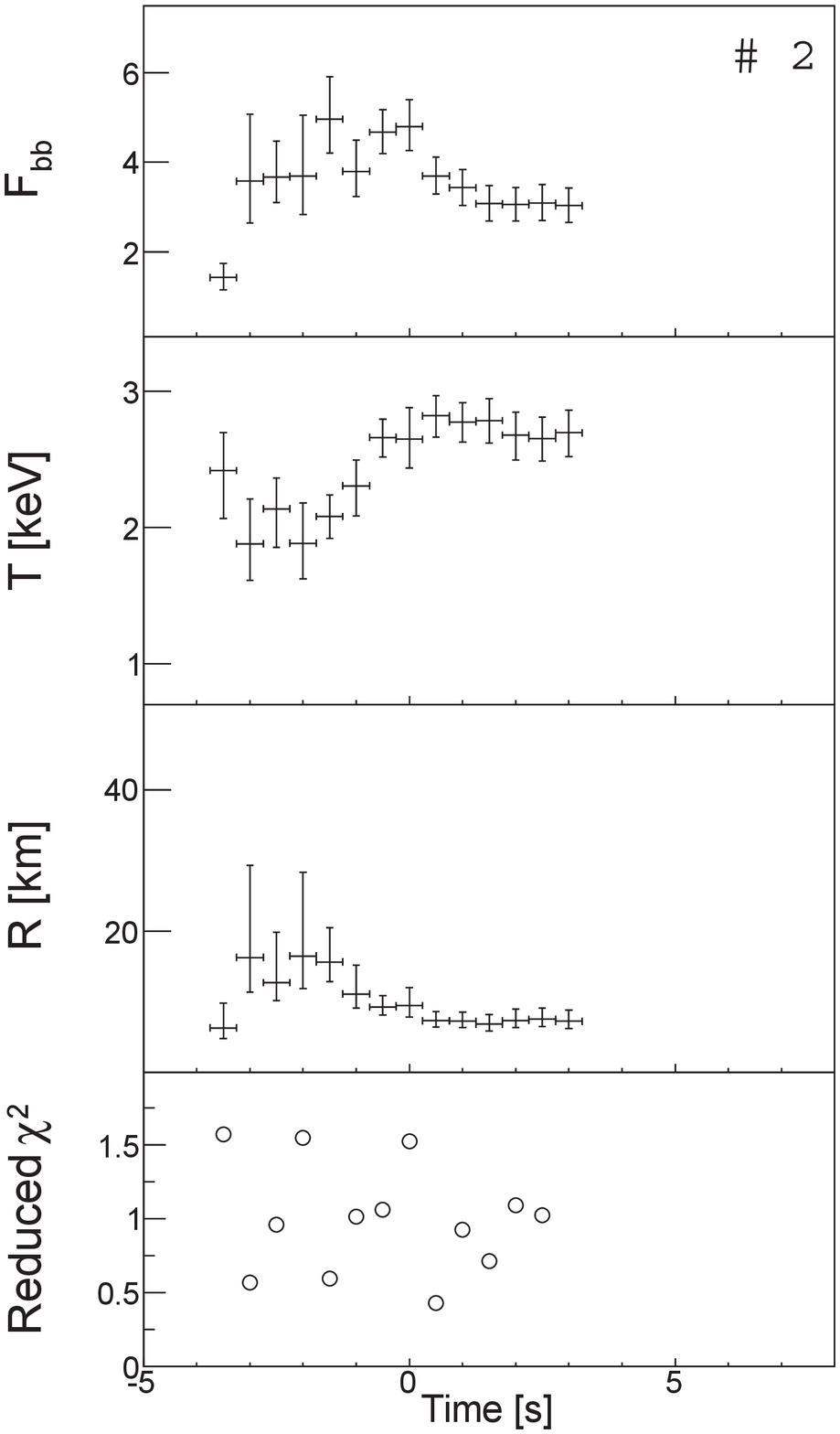}
\includegraphics[angle=0, scale=0.2]{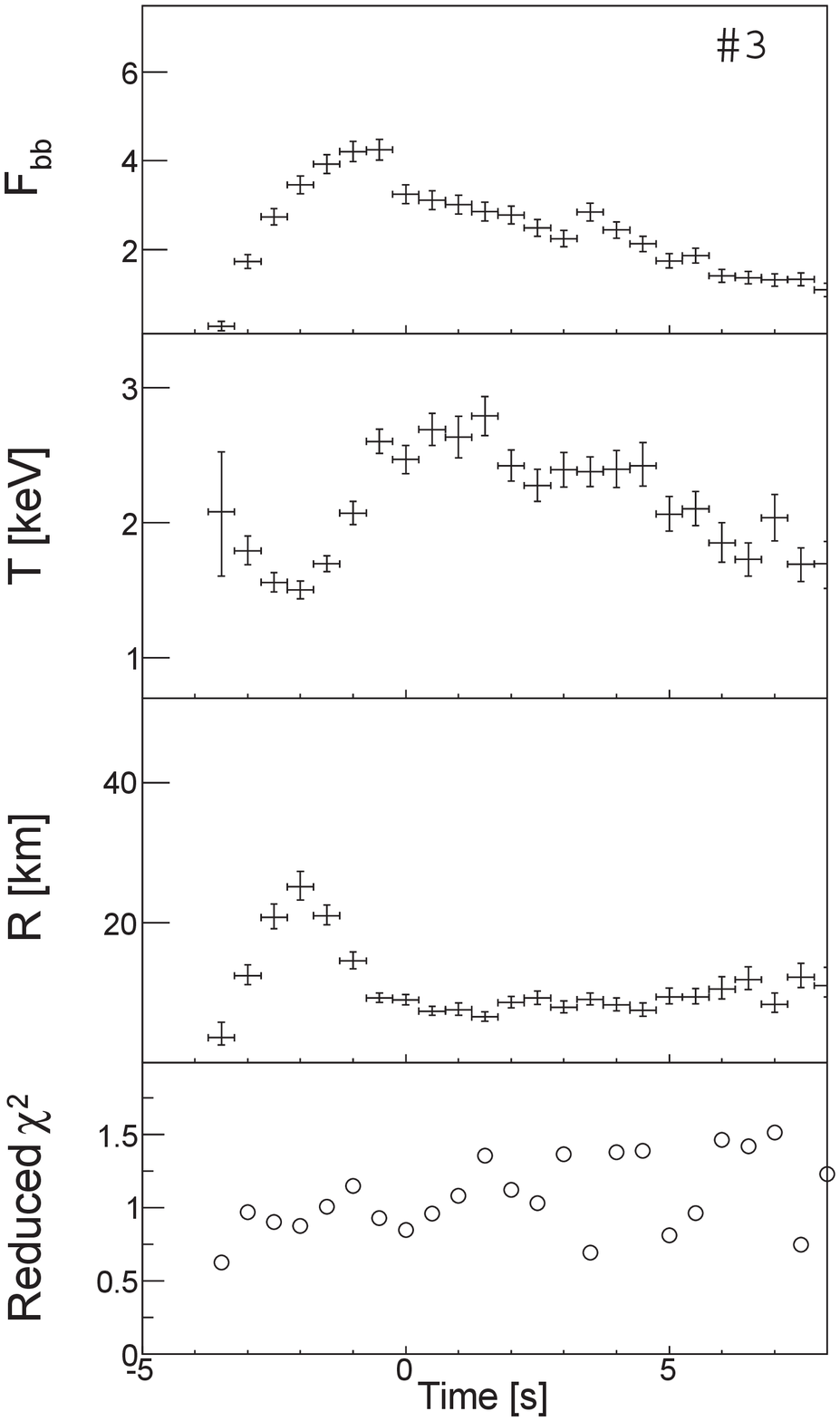}
\includegraphics[angle=0, scale=0.2]{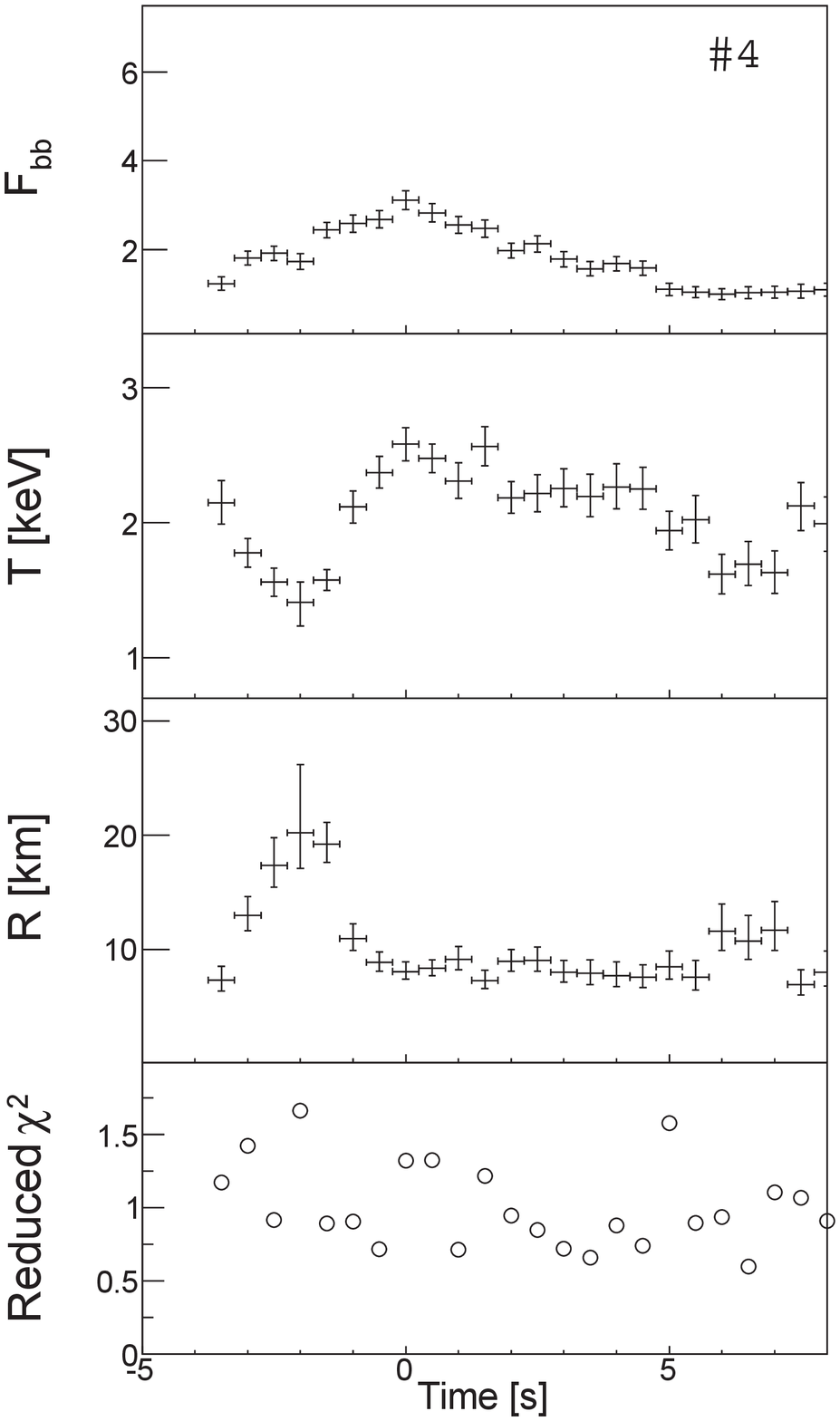}
\includegraphics[angle=0, scale=0.2]{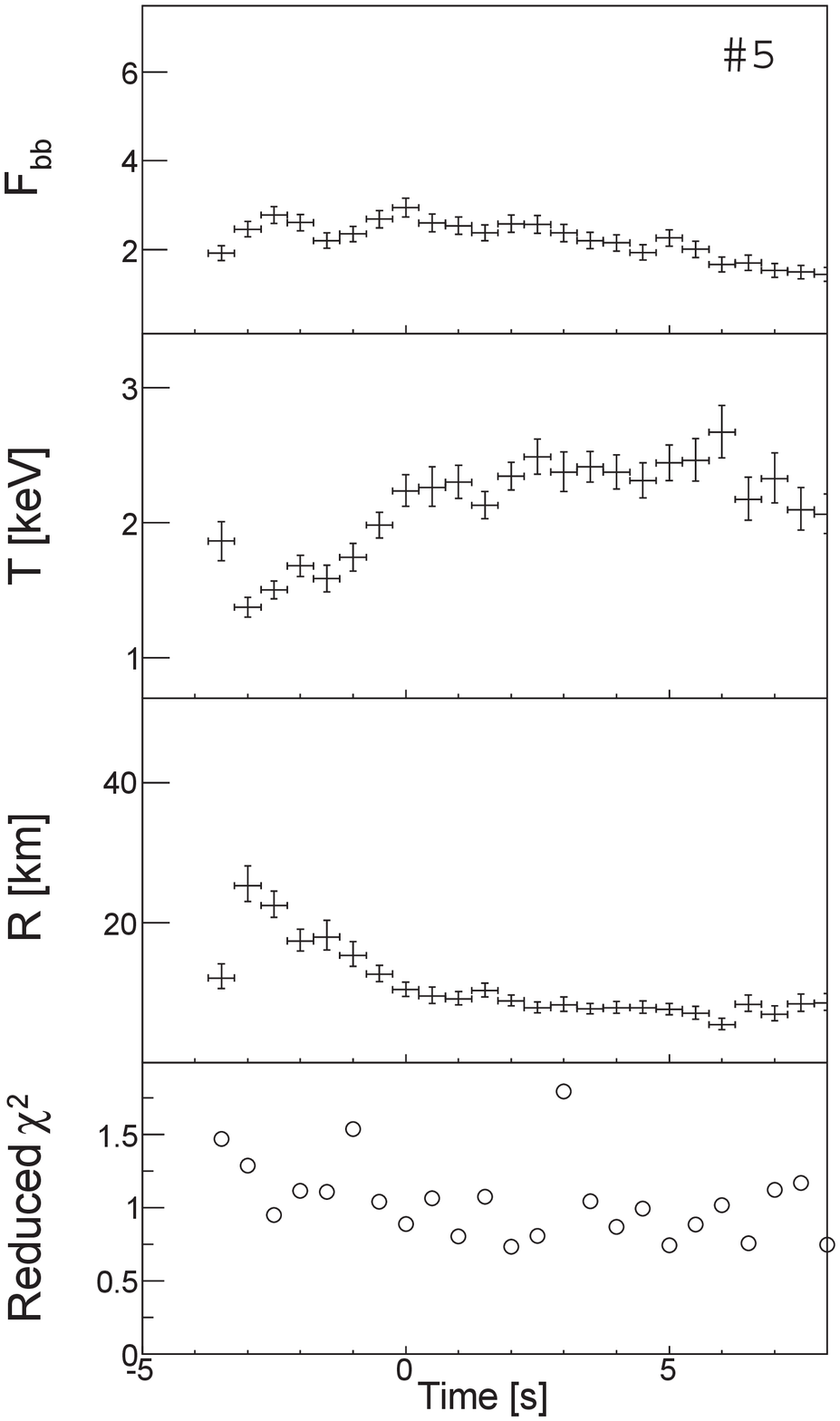}
\includegraphics[angle=0, scale=0.2]{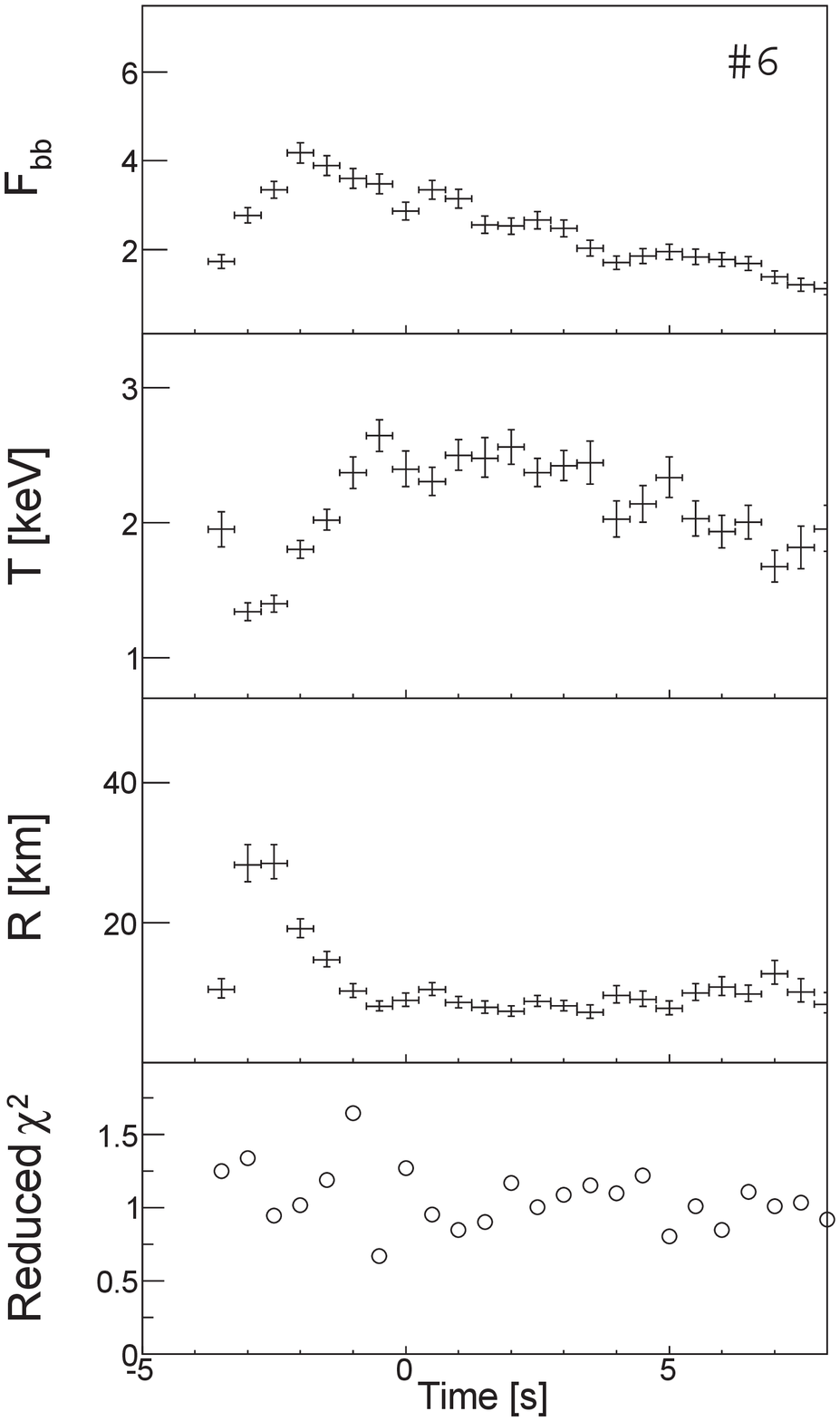}
\includegraphics[angle=0, scale=0.2]{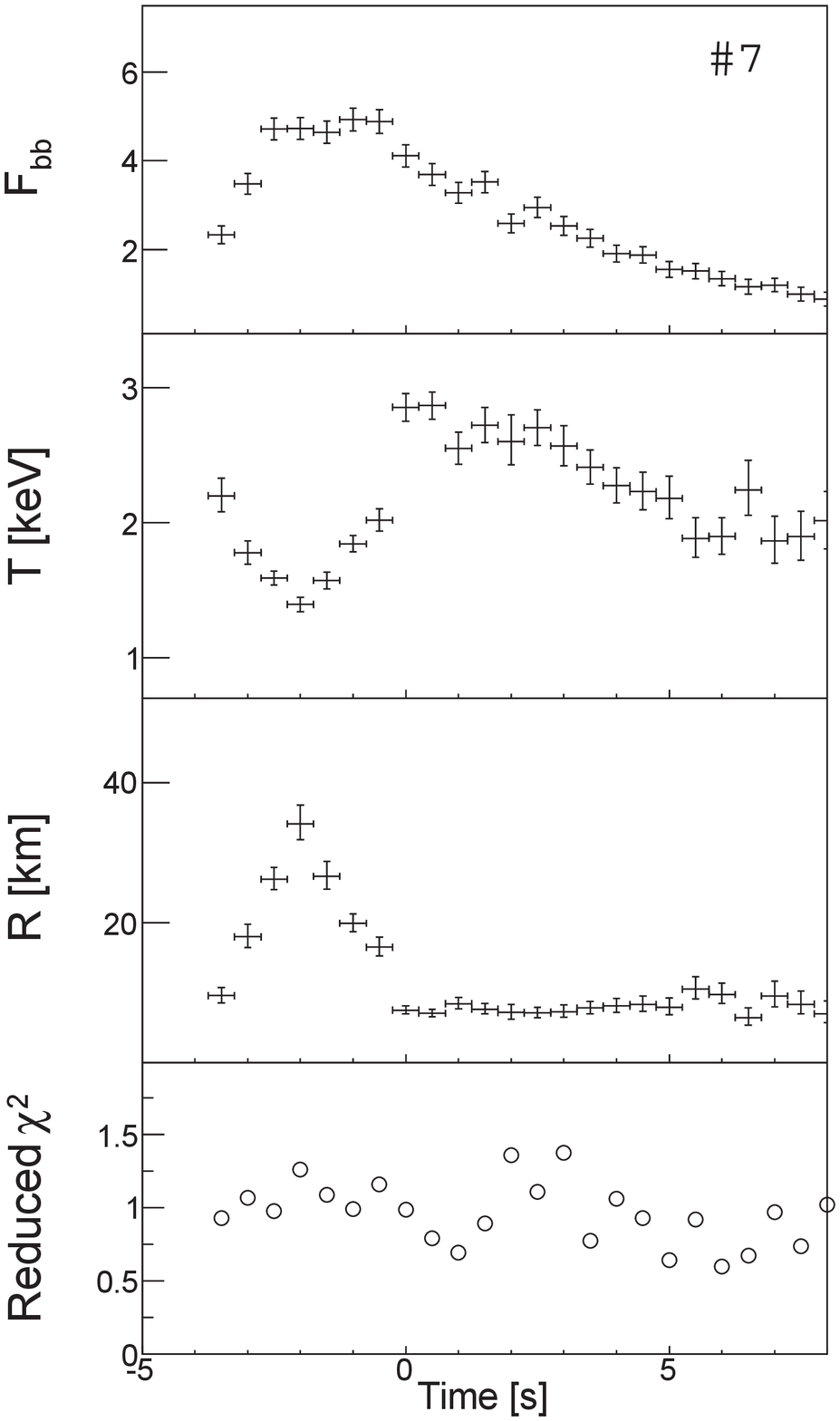}
\includegraphics[angle=0, scale=0.2]{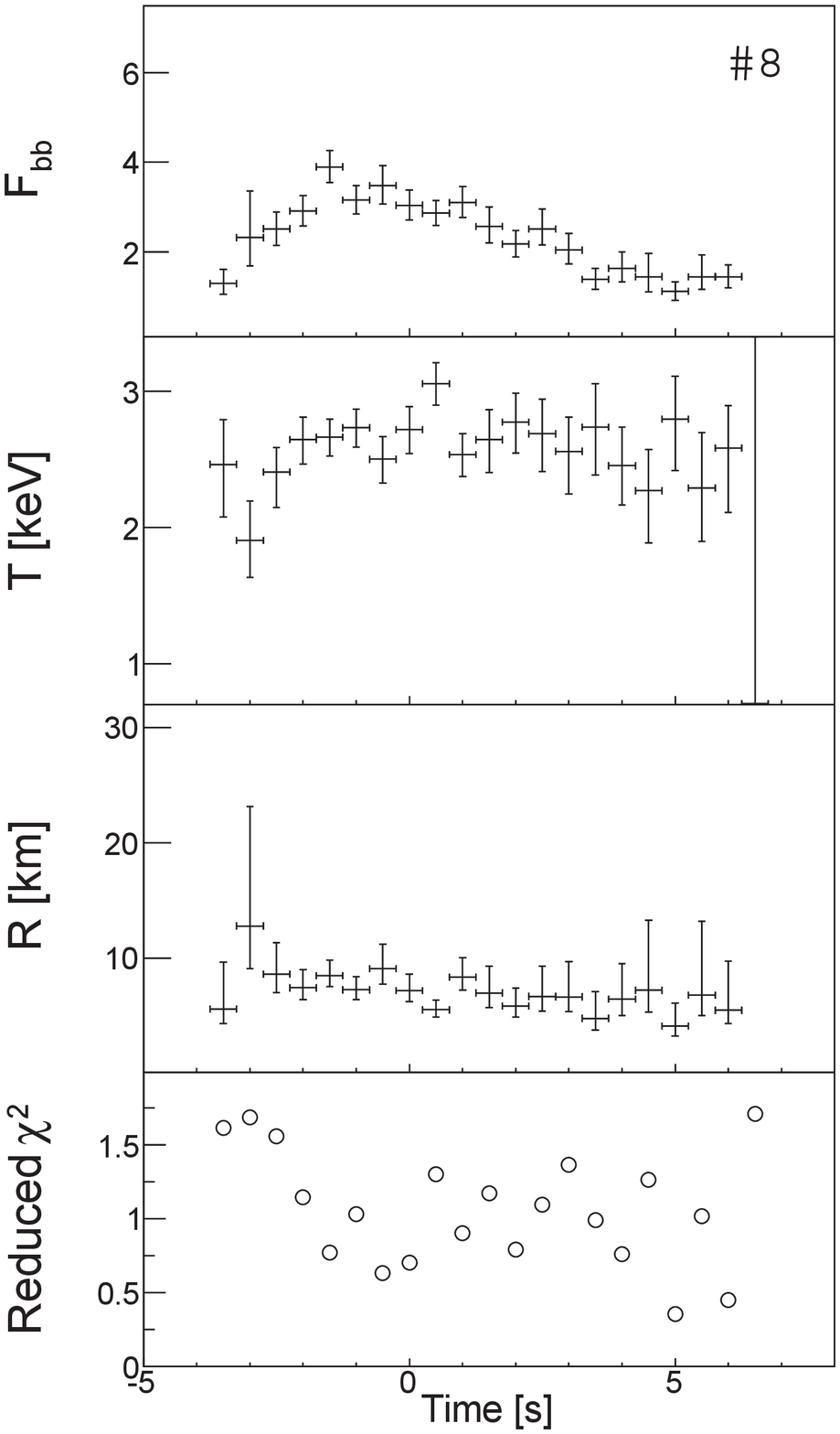}
\includegraphics[angle=0, scale=0.2]{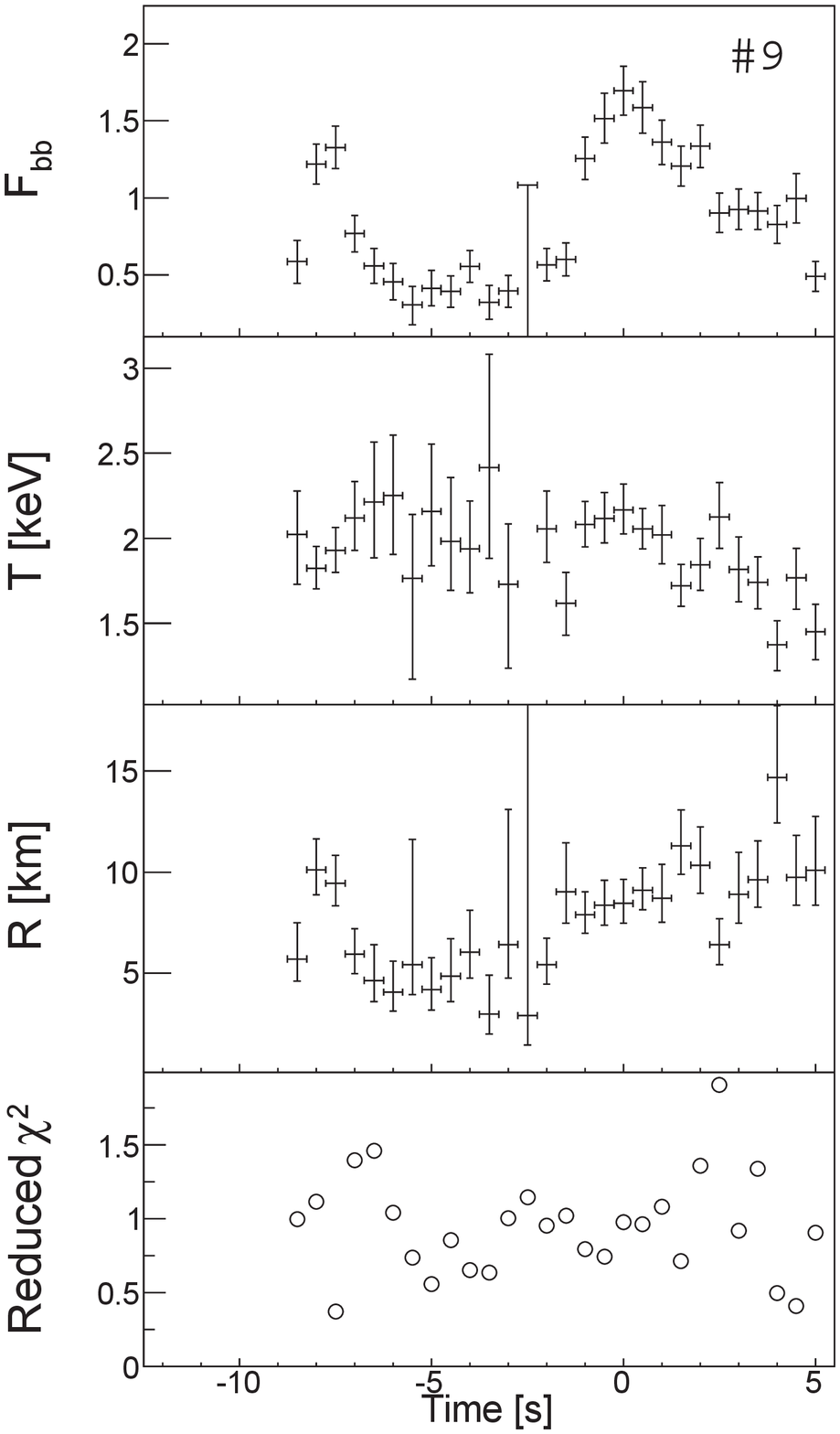}
\includegraphics[angle=0, scale=0.2]{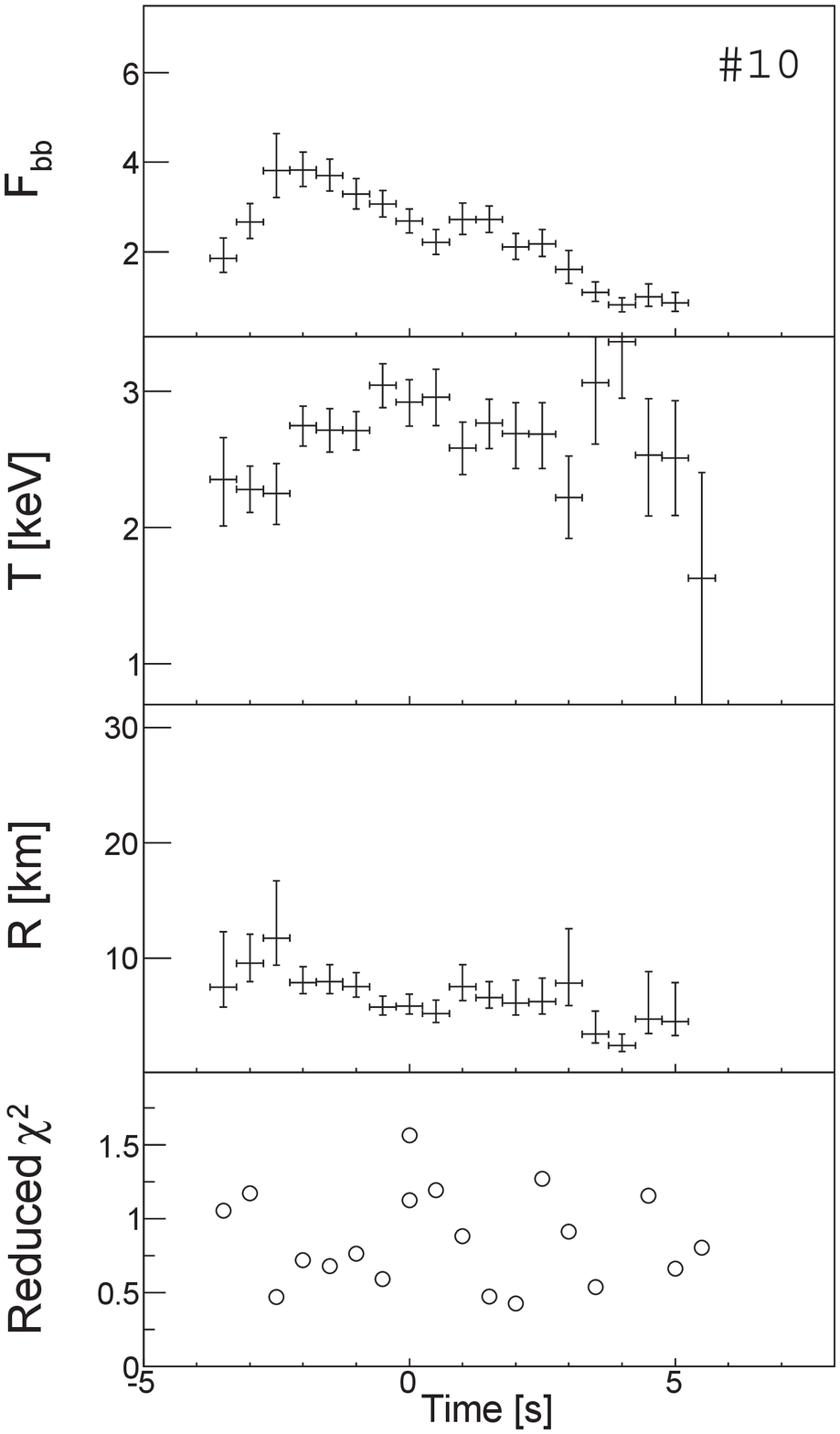}
 \caption{
 Results of the spectral fits of time-resolved spectra of the ten bursts detected from 4U~1730--22 during its 2021 and 2022 outbursts. All of the bursts except the  second-to-last burst show photospheric radius expansion.
  }
\label{fig_burst_fit_bb}
\end{figure}

We follow the classical approach to X-ray burst spectroscopy by subtracting the persistent spectrum
and fitting the net spectrum with an absorbed blackbody, as shown in Figure \ref{fig_burst_fit_bb}.
In the decay phase, such a spectral model generally results in  acceptable goodness-of-fit, with a mean reduced $\chi^{2}_{\upsilon}~\sim$ 1.0 (d.o.f. 20--60).
However, we note that  significant residuals are shown below 3 keV and above 10 keV, as shown in Figure \ref{fig_spec_residual}, especially for the  spectra in the PRE phase with the  reduced $\chi^{2}_{\upsilon}$ $>$ 1.5 (d.o.f. 60--80).

From the fitting results by the  absorbed blackbody, among the ten bursts, 9 bursts are PRE bursts with peak radii 12--40 km, peak temperatures $\sim$ 3 keV and peak bolometric fluxes 3--5$\times10^{-8}~{\rm erg}~{\rm cm}^{-2}~{\rm s}^{-1}$.
The model parameters of the bursts without LE detection, i.e., bursts \#1, \#2, \#8, and \#10,
show greater errors than that of the bursts with LE detection, which prevents us from adopting other models.
A similar situation exists in the faintest burst, burst \#9.

\subsubsection{Fit the spectra of bursts by the $f_{a}$ model }

To reduce the residuals, we first consider the $f_{a}$ model to fit the bright bursts which were detected simultaneously by LE, ME, and HE: burst \#3--burst \#7. Following \citet{Worpel2013} we then include an additional component for fitting the variable persistent emission.
We assume that during the burst the spectral shape of the persistent emission is unchanged, and only its normalization (known as the $f_{a}$ factor) is changeable.
As reported earlier by RXTE and NICER,
the $f_{a}$ model provides a better fit than the conventional one (absorbed blackbody).
We compare the above two models  using the F-test. 
In some cases, the $f_{a}$  model   significantly improves the fits with a p-value $\sim10^{-5}$.

\begin{figure}[t]
\centering
\includegraphics[angle=0, scale=0.4]{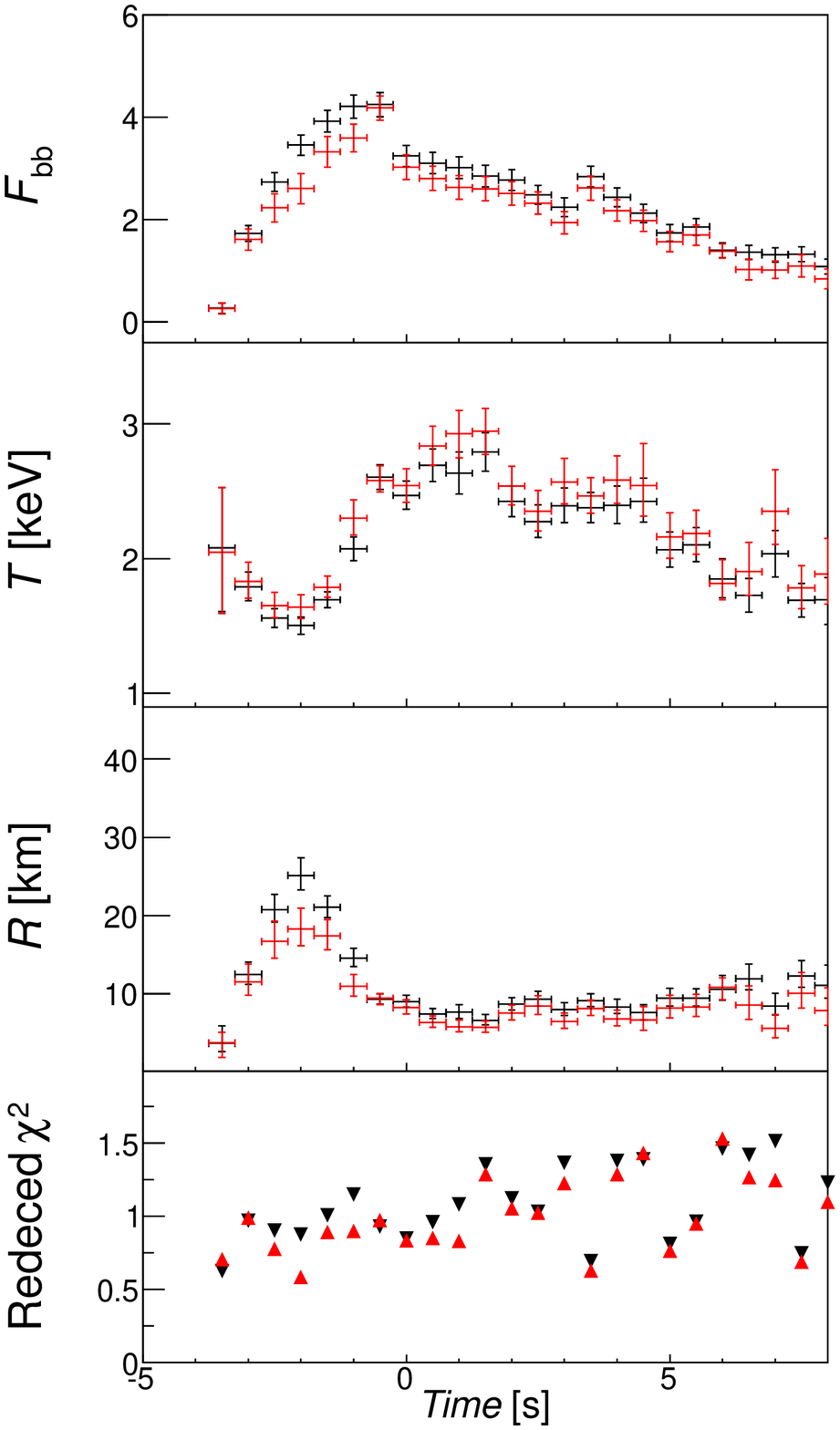}
\includegraphics[angle=0, scale=0.4]{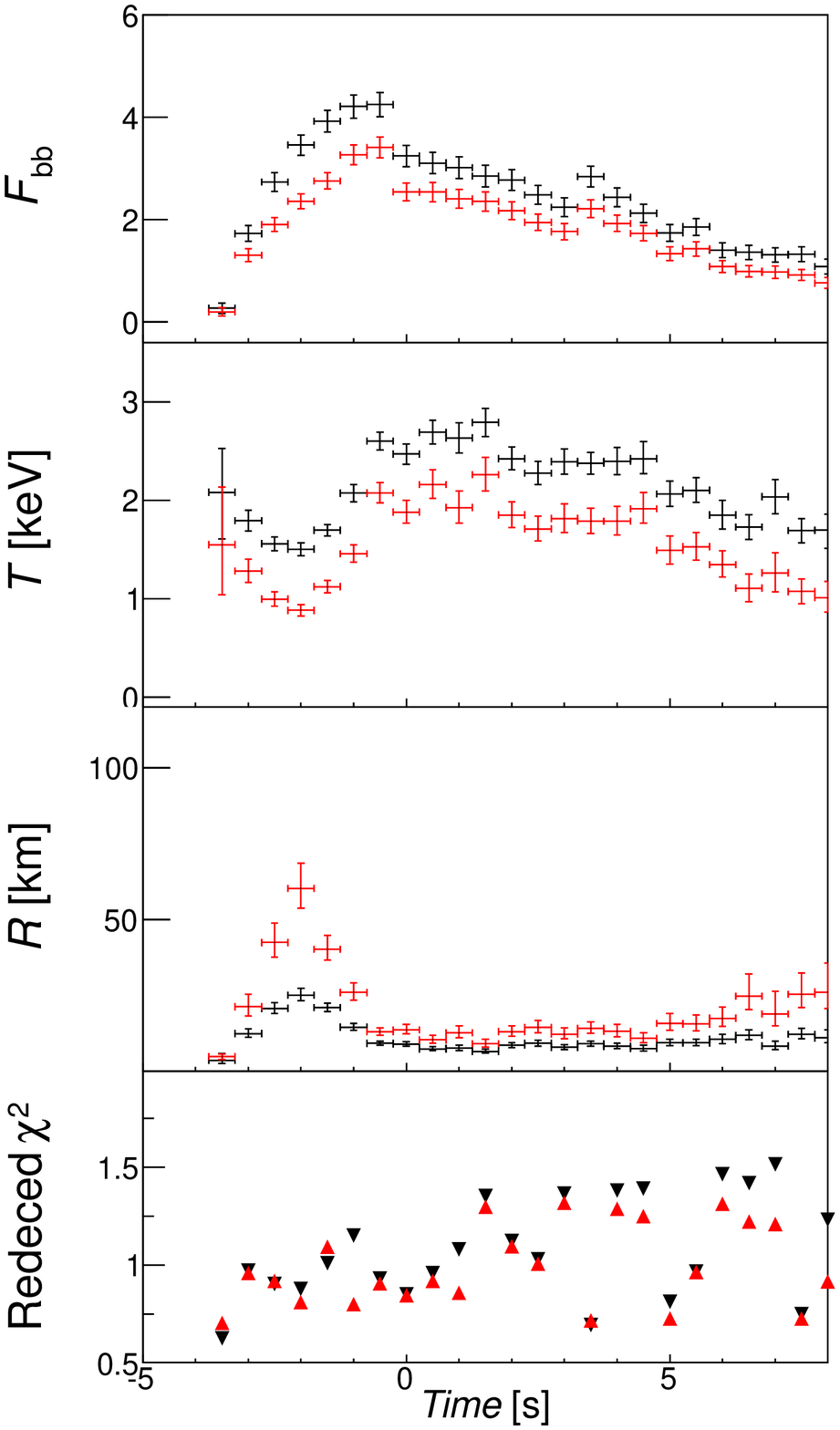}
  \caption{
Spectral fitting result of   burst \#3 with time bin 0.5 second  with  a pure blackbody (black), $f_{a}$ model (the left panel, red) and convolution thermal-Comptonization model (the right panle, red), 
include the time evolution of the blackbody bolometric flux $F_{\rm bb}$, the temperature $kT_{\rm bb}$, the observed radius $R$ of NS surface at 10 kpc, the goodness of fit $\chi_{v}^{2}$.
The bolometric flux of the blackbody model $F_{\rm bb}$ is in unit of $10^{-8}~{\rm erg/cm}^{2}/{\rm s}$.  }
\label{fig_fit_burst_P051400200102}
\end{figure}

\begin{figure}[t]
\centering
\includegraphics[angle=0, scale=0.4]{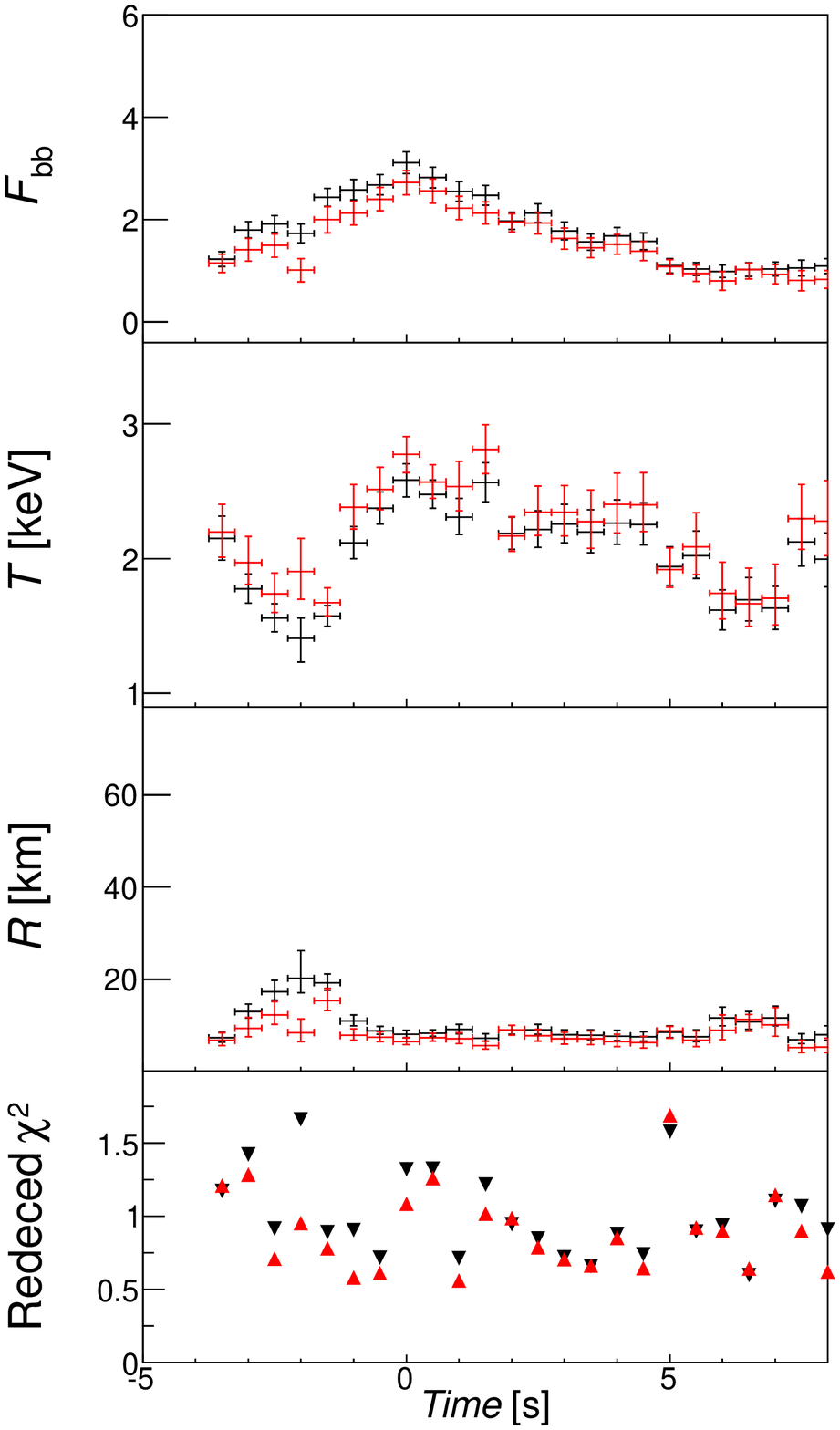}
\includegraphics[angle=0, scale=0.4]{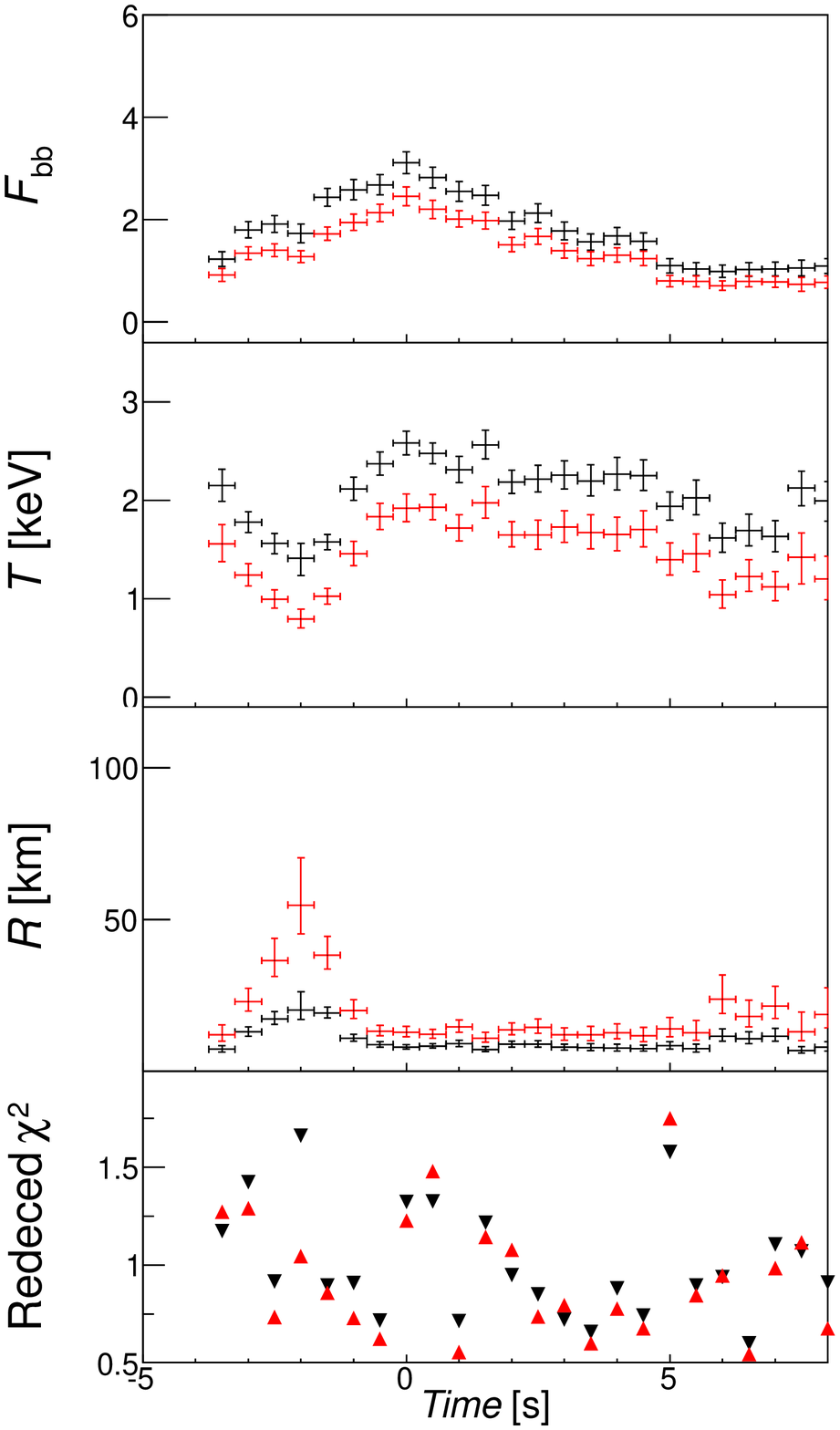}
  \caption{Same as Figure \ref{fig_fit_burst_P051400200102} for    burst \#4.
}
\label{fig_fit_burst_P051400200402}
\end{figure}

\begin{figure}[t]
\centering

\includegraphics[angle=0, scale=0.4]{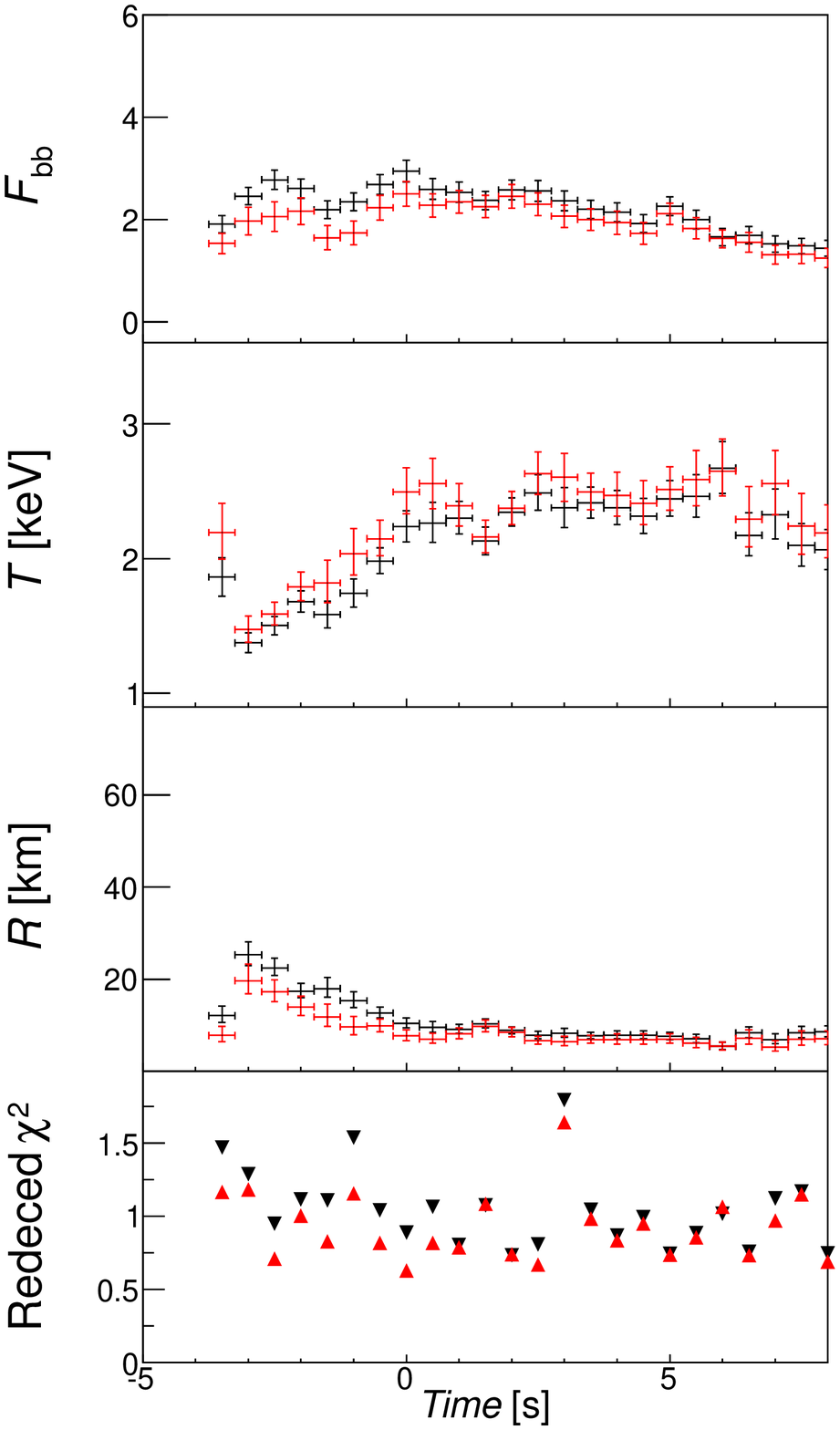}
\includegraphics[angle=0, scale=0.4]{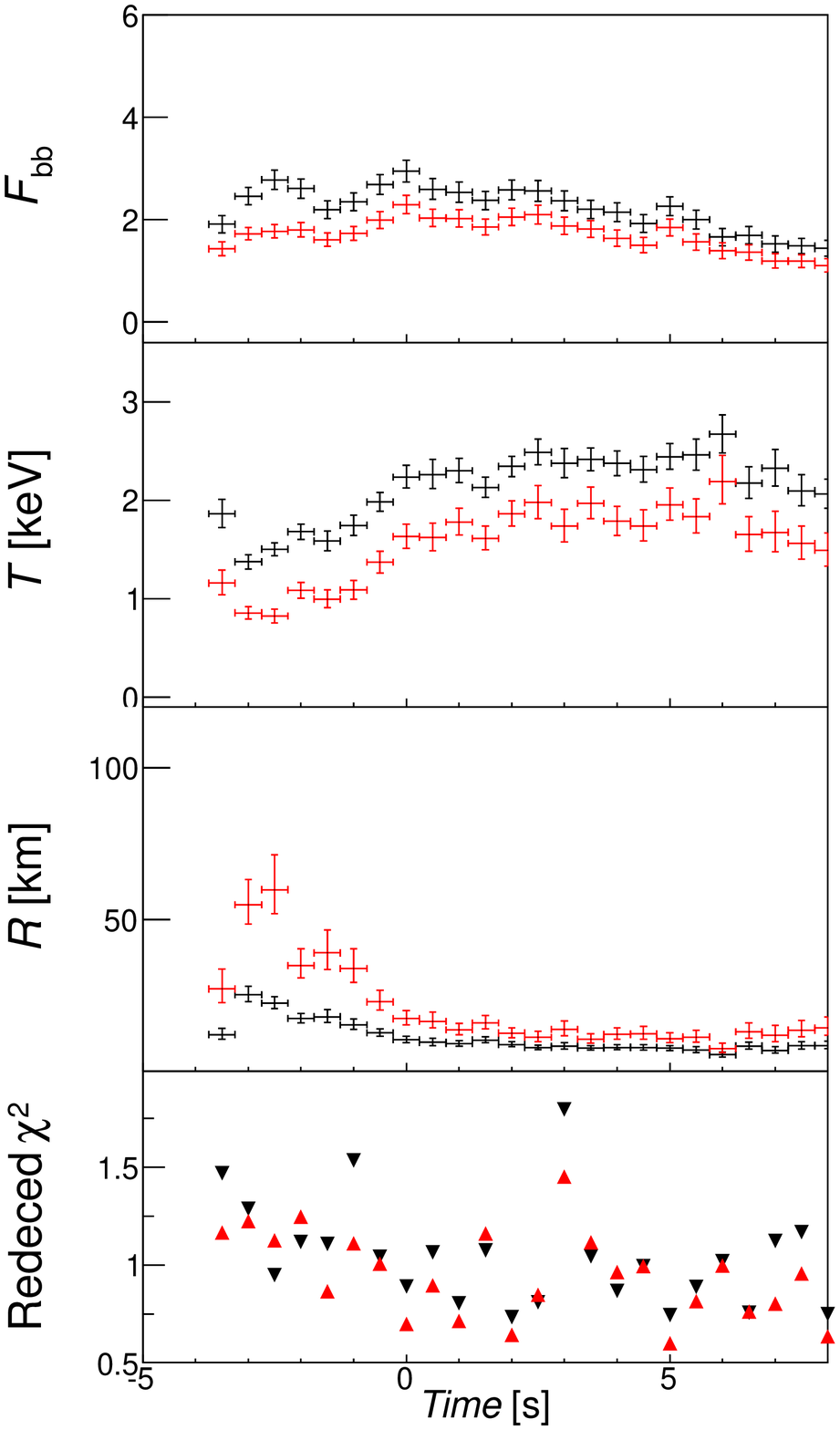}

 \caption{Same as Figure \ref{fig_fit_burst_P051400200102} for    burst \#5.
}
\label{fig_fit_burst_P051400200601}
\end{figure}

\begin{figure}[t]
\centering

\includegraphics[angle=0, scale=0.4]{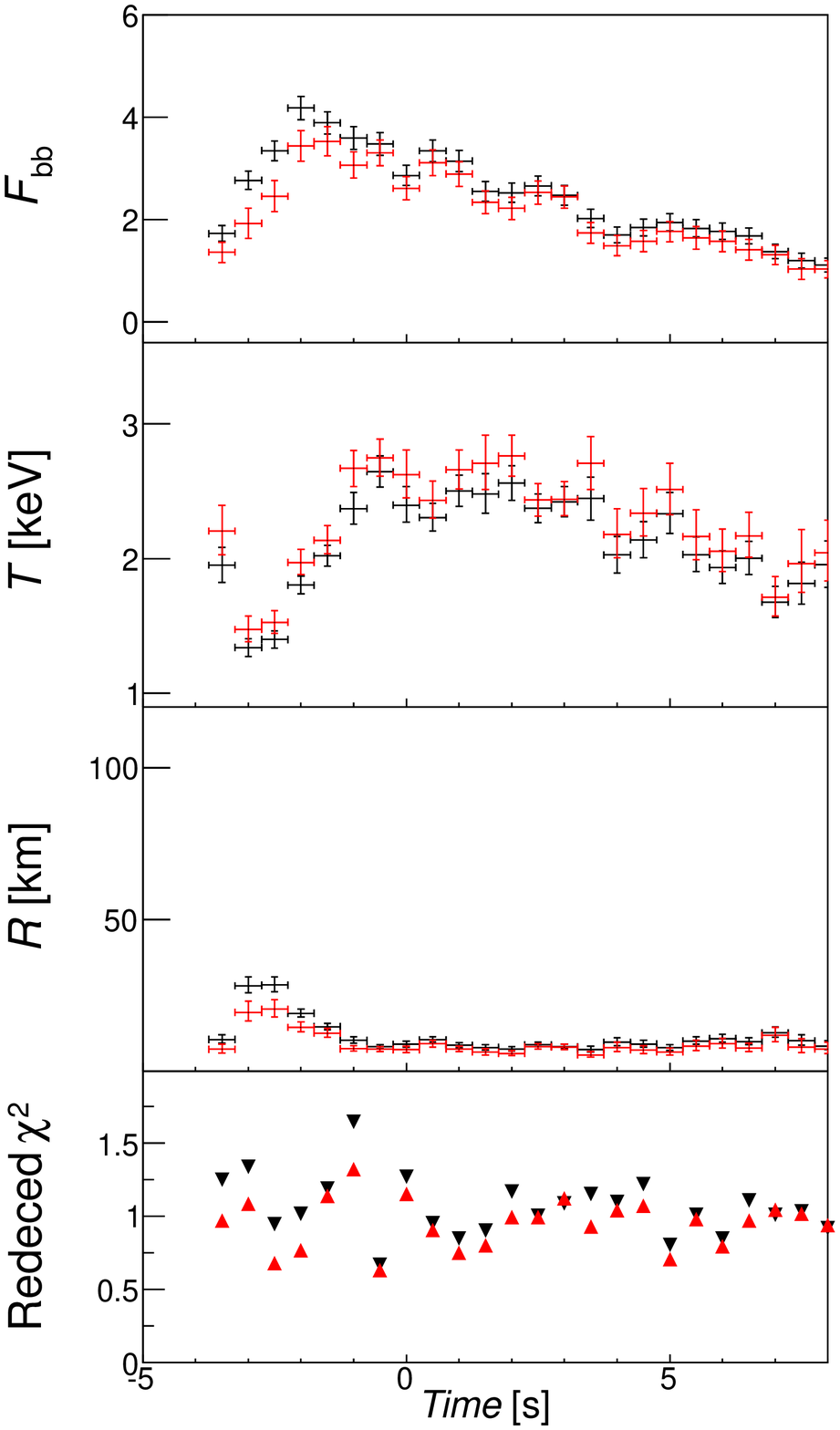}
\includegraphics[angle=0, scale=0.4]{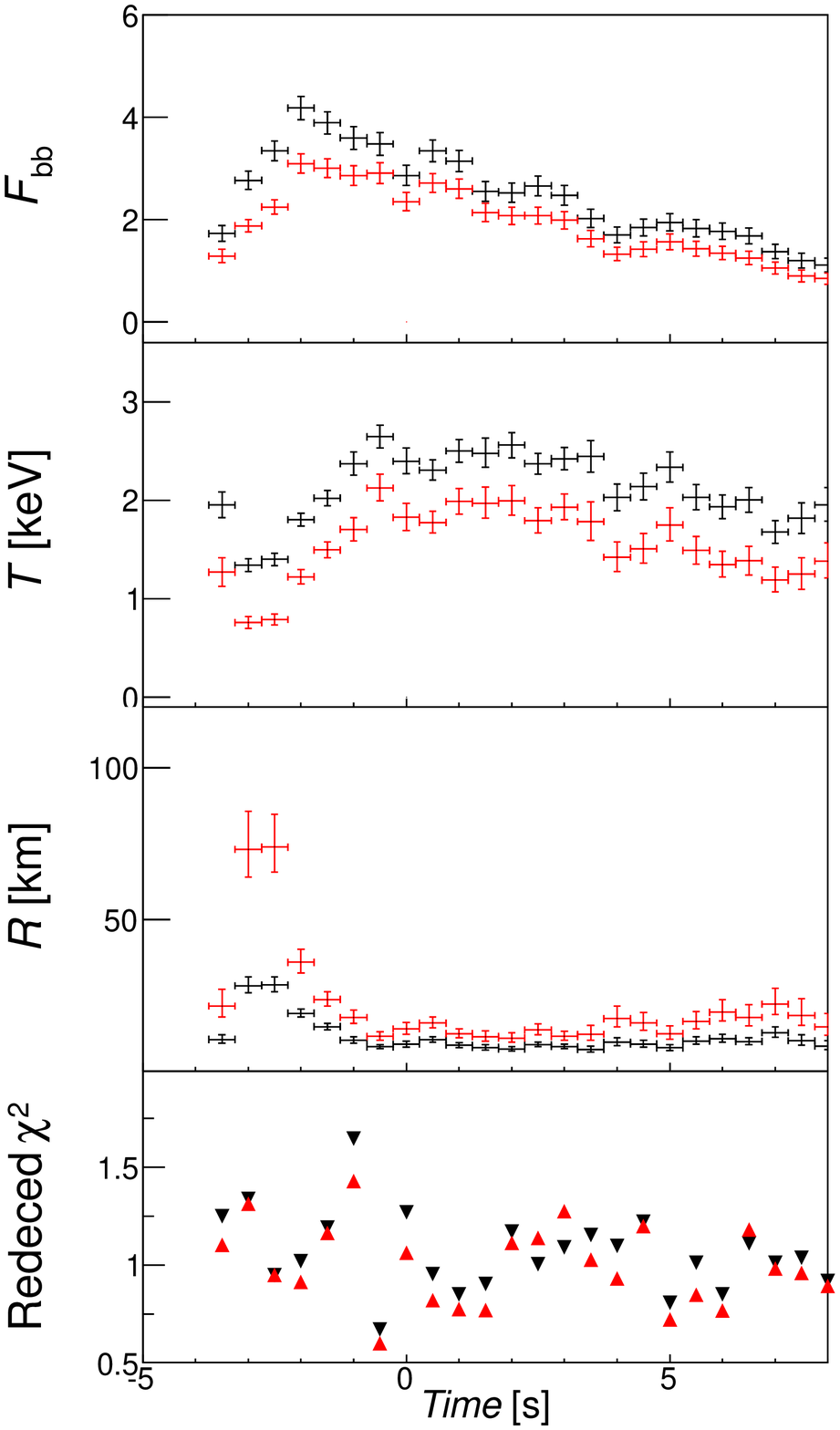}

 \caption{Same as Figure \ref{fig_fit_burst_P051400200102} for    burst \#6.
}
\label{fig_fit_burst_P051400200701}
\end{figure}

\begin{figure}[t]
\centering
 \includegraphics[angle=0, scale=0.4]{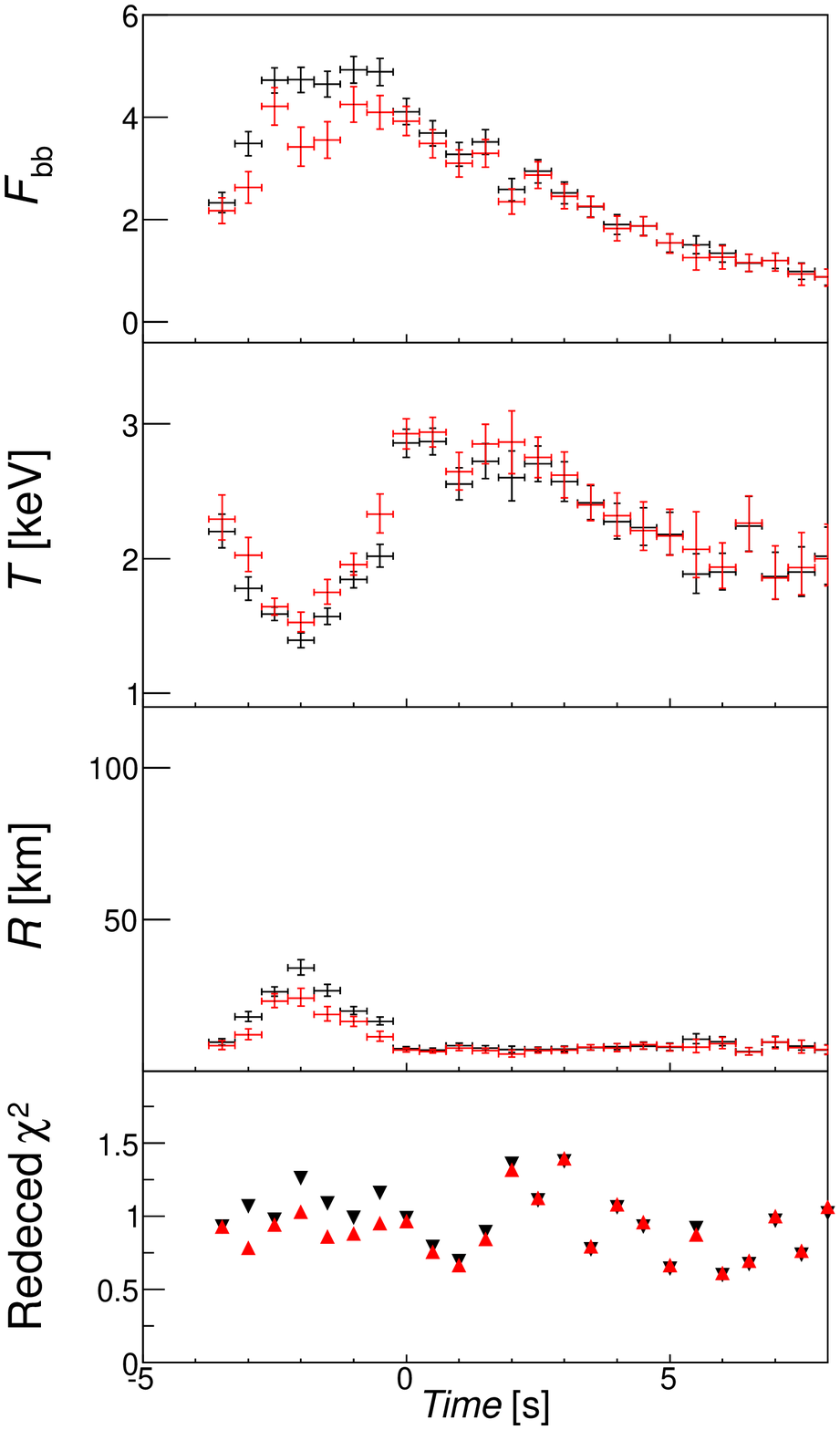}
\includegraphics[angle=0, scale=0.4]{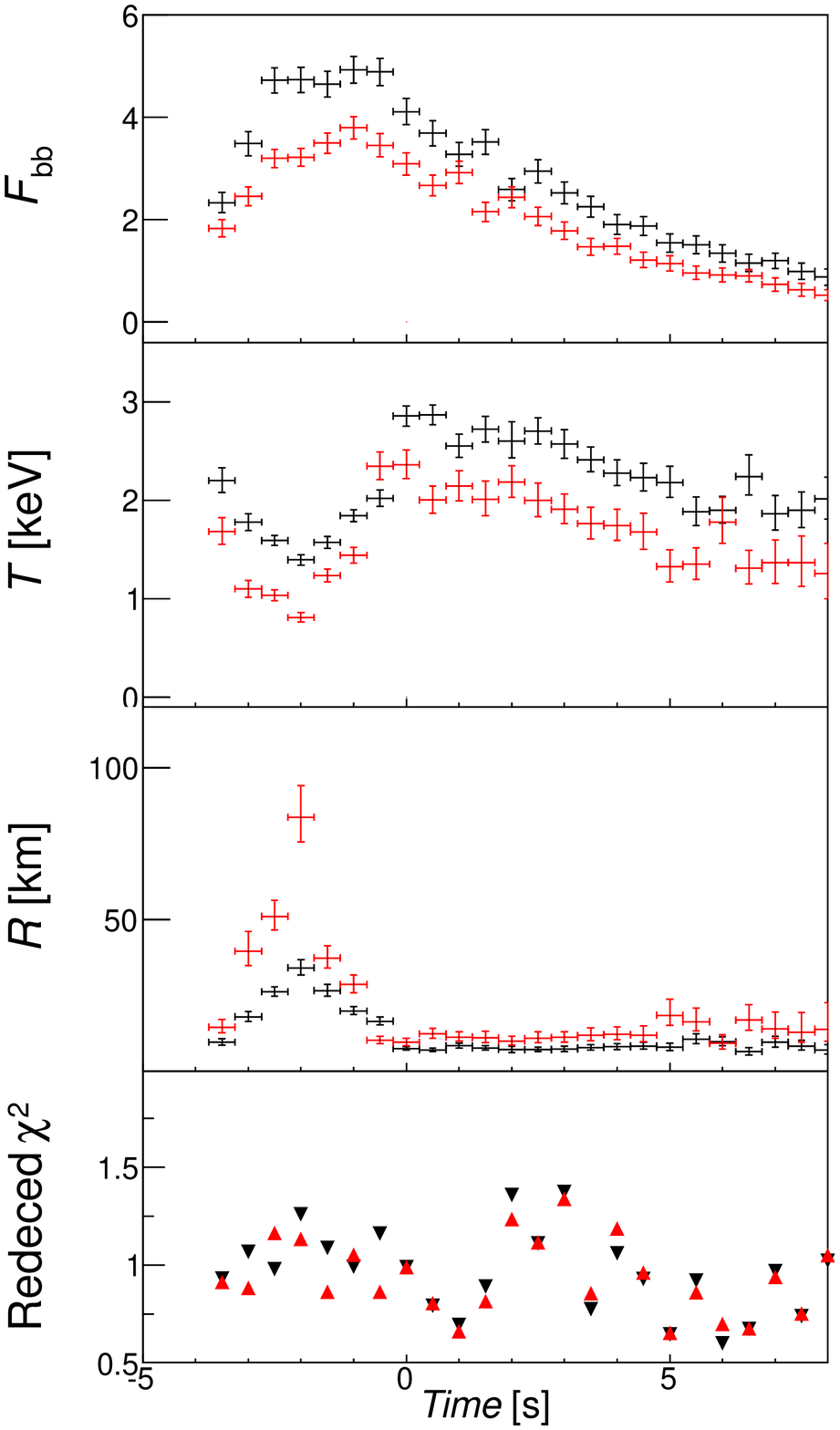}
 \caption{ Same as Figure \ref{fig_fit_burst_P051400200102} for    burst \#7.
}
\label{fig_fit_burst_P051400200801}
\end{figure}

\begin{figure}[t]
\centering
      \includegraphics[angle=0, scale=0.30]{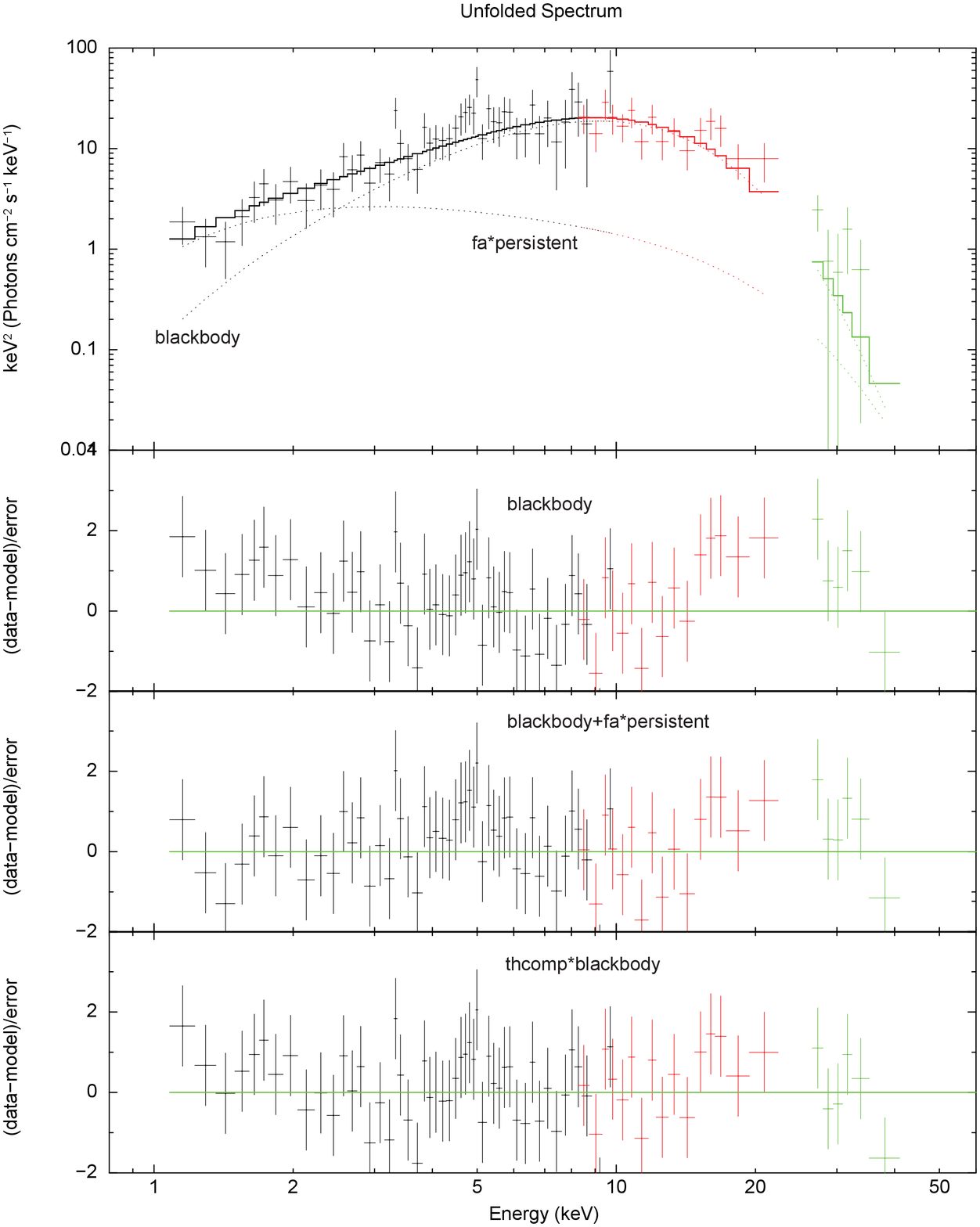}
 \caption{Top panel: the spectral fits   results by  LE (black), ME (red) and HE (green) when the burst \#7 reaches the touch-down time by $f_{a}$ model, the blackbody model and enhancement of the persistent emission are labeled.
 The three panels below: residuals of spectral fits results by
 an absorbed black-body  model (the 2nd panel), $f_{a}$ model (the 3rd panel) and convolution thermal-Comptonization model (the bottom panel).
 }
\label{fig_spec_residual}
\end{figure}

As shown in the left panels of Figures \ref{fig_fit_burst_P051400200102}, \ref{fig_fit_burst_P051400200402}, \ref{fig_fit_burst_P051400200601}, \ref{fig_fit_burst_P051400200701}, and \ref{fig_fit_burst_P051400200801}, the spectral fitting results from these two models have differences mainly around the PRE phase.  By considering an additional factor $f_{a}$, the burst blackbody flux tends to slightly decrease, and the temperature becomes higher but the radius shrinks.
Using the average flux of the touch-down time of the five bursts  $2.92\pm0.11\times10^{-8}~{\rm erg/cm}^{2}/{\rm s}$, and assuming the empirical Eddington luminosity of $3.8\times10^{38}$ erg/s \citep{Kuulkers2003}, we derive  the source distance of 10.4 kpc.
For simplicity, we use a distance of 10 kpc to calculate the luminosity and blackbody radius.




The $f_{a}$ factor reaches a maximum of $6\pm1$ when the radius reaches its peak.  
During the PRE phase, the radius is up to $\sim 30$ km, which is four times larger than the radius measured at touch-down time  $\sim$ 8 km (assuming a distance of 10 kpc).
This is typical of a moderate photospheric expansion.

\subsubsection{Fit the spectra of bursts by the  convolution  thermal Comptonization model}

Since the burst photons could also be affected by the corona/boundary-layer, we thus check if the model used in the persistent emission could be the same as the burst emission.
By taking the pre-burst emission as background emission, the burst spectra are fitted by the model tbabs*thcomp*bb, in which the thcomp parameters are fixed at the persistent emission fit results.  Thus the convolution  thermal Comptonization model (with an input seed photon spectrum of  blackbody)  has the same d.o.f as the canonical blackbody model, and a more d.o.f. than the $f_{a}$ model.  The bb and thcomp represent the burst emission from the NS photosphere and  a corona/boundary-layer influence on the burst emission.
This model allows us to evaluate  the contribution from both the up-scattered by the corona/boundayr-layer and direct  photons from the NS surface.

As shown in the right panels of Figures \ref{fig_fit_burst_P051400200102}, \ref{fig_fit_burst_P051400200402}, \ref{fig_fit_burst_P051400200601}, \ref{fig_fit_burst_P051400200701}, and \ref{fig_fit_burst_P051400200801},
in the PPE phase, this model provides the best fit and
yields physically acceptable spectral parameters; the obtained best-fit parameters are given in the right panels. We find that this convolved  thermal-Comptonization model provides   equally good results as the $f_{a}$ model.
As mentioned above, the free/unfixed parameters include the blackbody temperature  $kT_{\rm bb}$ and the normalization $N_{\rm bb}$. The trend of the parameters is similar to the $f_{a}$ model, but with a greater change. Compared to the $f_{a}$ model results, the maximum radius $R_{\rm bb}$ is up to $83.7_{-8.2}^{+10.4}$ km, the minimum temperature  $kT_{\rm bb}$ is lowered to $0.81\pm0.05$ keV.
Using the $f_{a}$ model,
the source distance is estimated as 11.2 kpc with  the average flux of $2.51\pm0.08\times10^{-8}~{\rm erg/cm}^{2}/{\rm s}$ derived at the touch-down time of the five bursts, from the convolved  thermal-Comptonization model.
Other scenarios, i.e.,  burst reflection by the disk, NS atmosphere  model carbatm/hatm \citep{Suleimanov2011,Suleimanov2012,Suleimanov2018} in Xspec, are also tried to fit the burst spectra, as we did in \citet{Chen2019}.
However, neither could  alleviate the residuals at soft X-ray and hard X-ray bands simultaneously.

\subsubsection{No cooling between the sub-bursts of the double-peaked burst}

As shown in Figure \ref{fig_burst_fit_bb}, for the first sub-burst of   the double-peaked burst,  the temperature and the radius of the blackbody reach $1.9\pm0.1$ keV and  $9.5\pm1.4$ km. After that, the radius drops but the temperature stays at a high value: we average the eight data points (4 seconds) during the flux dip, and get an average temperature  of  $1.9\pm0.1$ keV and an average radius of $5.5\pm0.6$ km.
For the second sub-burst, it reaches  peak flux  up to 30 percent   brighter than the first one. 

\subsection{Rising bolometric flux during the PRE phase}

\begin{figure}[t]
\centering
    \includegraphics[angle=0, scale=0.20]{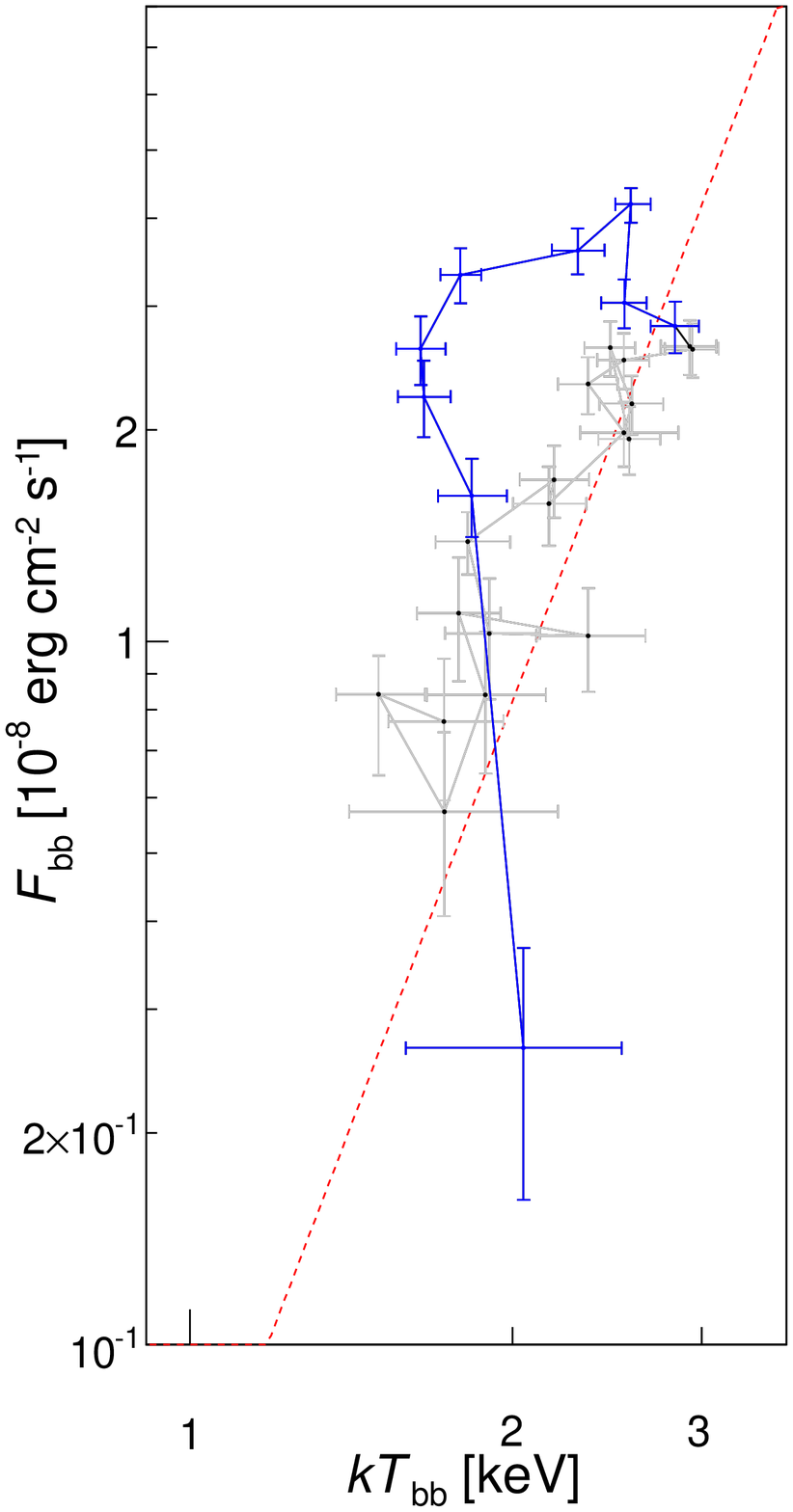}
    \includegraphics[angle=0, scale=0.20]{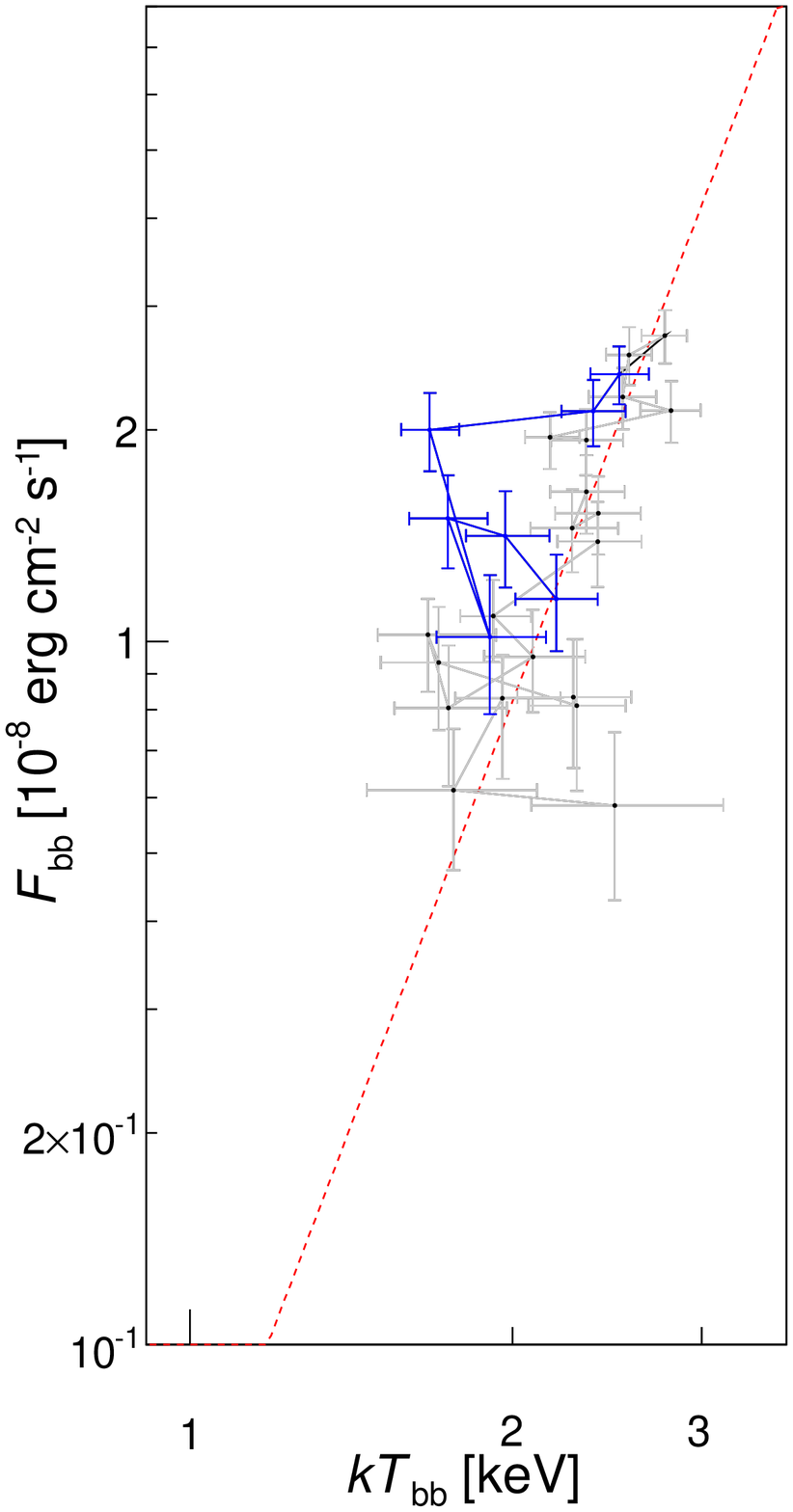}
    \includegraphics[angle=0, scale=0.20]{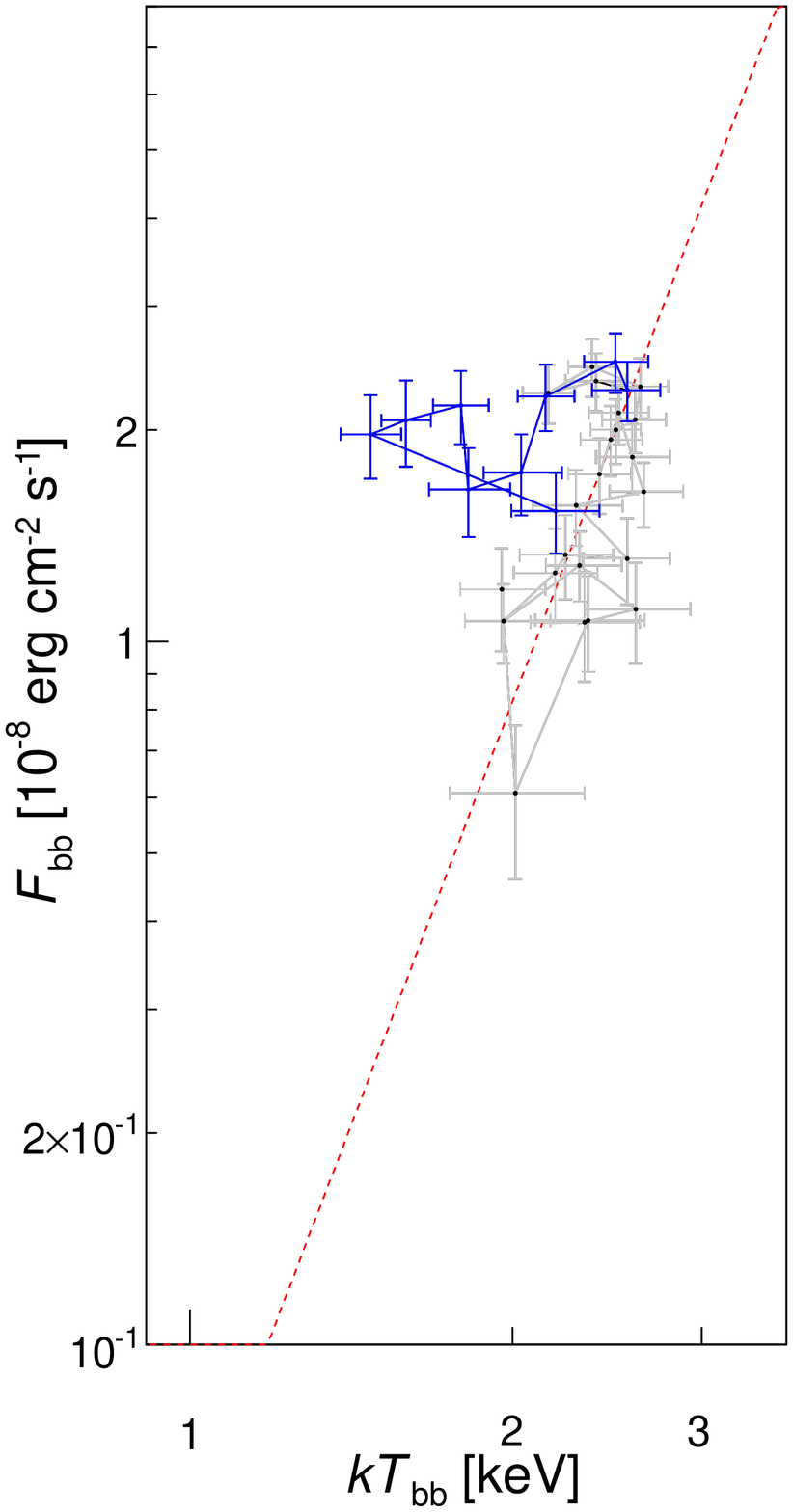}
    \includegraphics[angle=0, scale=0.20]{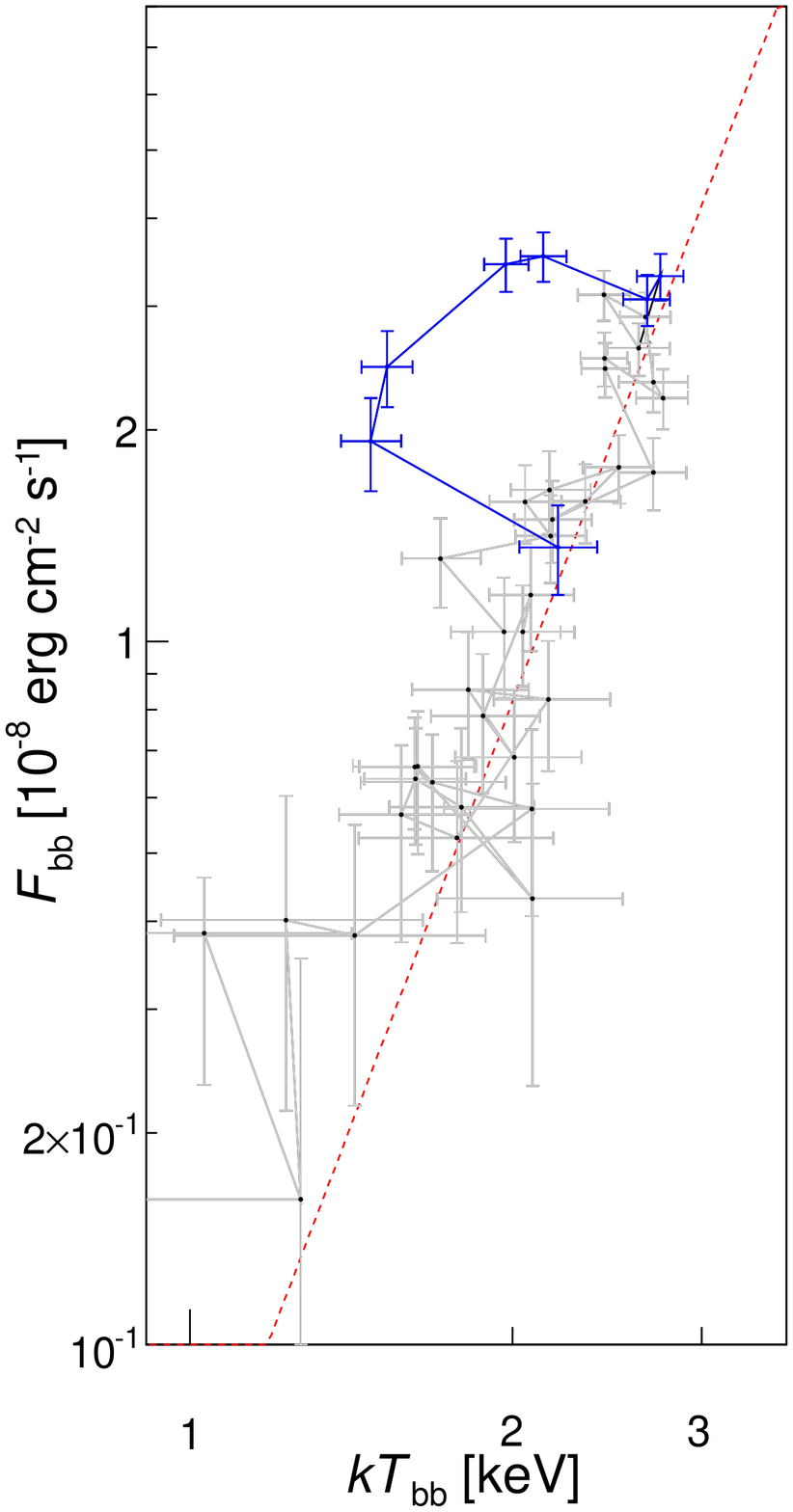}
    \includegraphics[angle=0, scale=0.20]{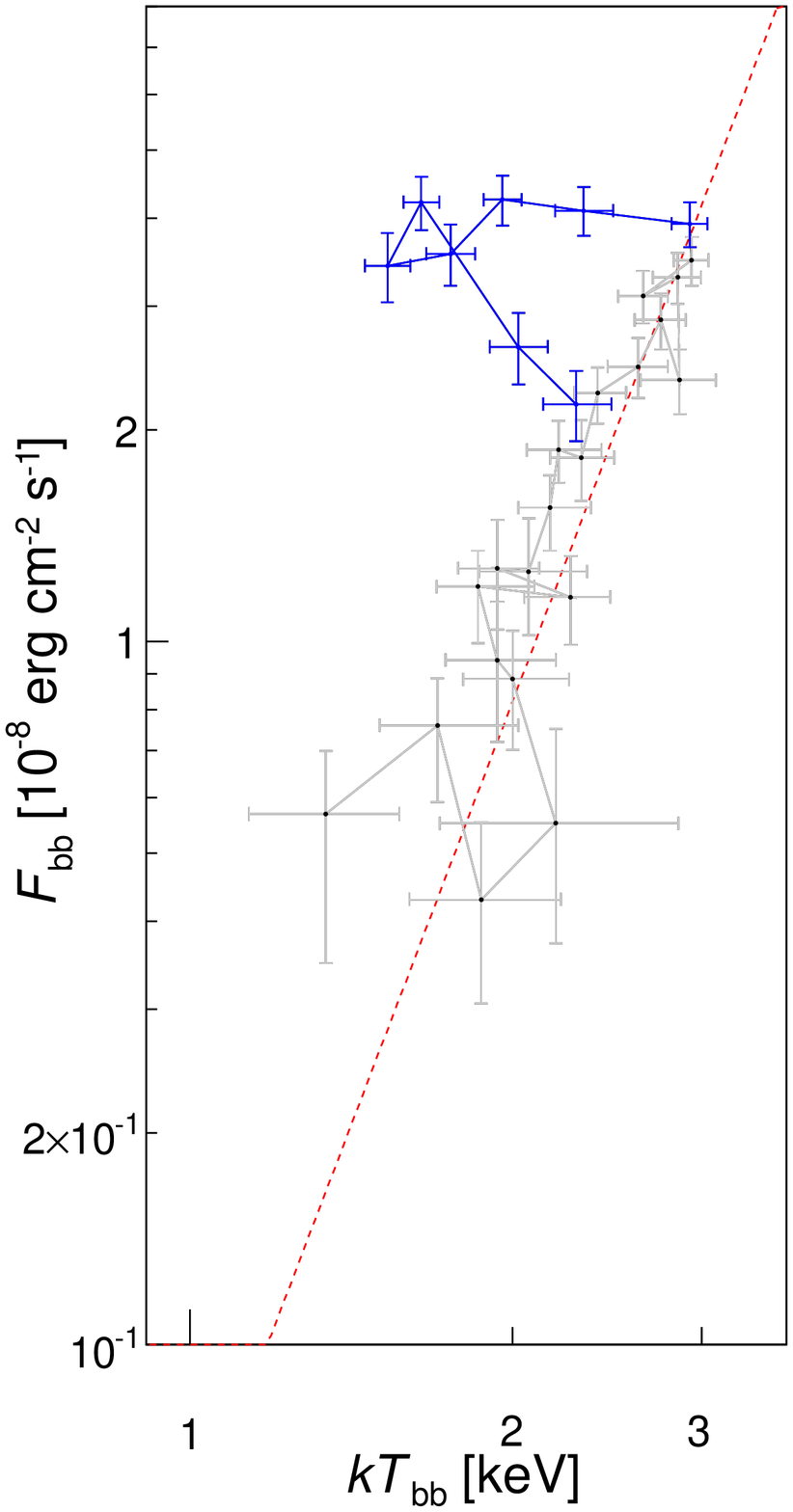}
        \caption{
        Burst flux vs. blackbody temperature of   bursts \#3--\#7. The blue points and gray points indicate the data points before and after the touch-down times. The dashed lines correspond to $R_{\rm BB}$=6.9 km under the distance of 10 kpc.
 }
\label{fig_t_f_6.9km}
\end{figure}

In Figure \ref{fig_t_f_6.9km}, we explore the relation between the bolometric flux, $F_{\rm bb}$, and the blackbody temperature, $kT_{\rm bb}$, using the parameters derived from the $f_{a}$ model for the five bursts. If the whole NS surface  emits as a single-temperature blackbody and a constant color correction factor, the burst flux $F$ should scale as $T_{\rm bb}^{4}$ in the flux-temperature diagram, and the slope represents the emitting area in the double logarithmic coordinates \citep{Guver2012}.
The diagram for the convolution  thermal Comptonization model is not given, since the trend is very similar with the $f_{a}$ model.
The diagonal line in the plot represents the line of constant radius, $R_{\rm bb}$=6.9 km,  assuming a distance $d$=10 kpc to the source, which is derived from the fitting of the decay phase of the bright burst (burst \#7) in the diagram of $F_{\rm bb}$ vs $T_{\rm bb}$ by a model $F_{\rm bb}\propto~R_{\rm bb}^{2}T_{\rm bb}^{4}$.

From the diagram, in the decay phase (gray points in Figure \ref{fig_t_f_6.9km}), it is apparent that the bursts follow the expected relation $F_{\rm bb}\propto~R_{\rm bb}^{2}T_{\rm bb}^{4}$.
In the PRE phase, i.e., the  photospheric radius larger than the NS radius  (blue points in Figure \ref{fig_t_f_6.9km}), the bursts depart from the $F_{\rm bb}\propto~R_{\rm bb}^{2}T_{\rm bb}^{4}$ relation and locate at the left of the line of $R_{\rm bb}$=6.9 km, which indicates   larger radii.
There are two  junctions between the blue points and the red line: the upper one   corresponds to  the touch-down time, and the lower one
 corresponds to the time when the photosphere  is just lifted from the NS surface.
We notice that the fluxes of the two junctions are different, i.e., the upper one is at least twice bright as the lower one, e.g. the two fluxes for burst \#3 are $0.7\times10^{-8}~{\rm erg/cm}^{2}/{\rm s}$ and $2\times10^{-8}~{\rm erg/cm}^{2}/{\rm s}$, which should be the same value since both of them are the values of the Eddington limit.




\section{Discussion}
In this work, we have presented a spectral analysis of ten bursts and persistent emission from 4U~1730--22 in its 2021 and 2022 outbursts observed by NICER and Insight-HXMT.
For the persistent emission in the outburst of 2022,
the joint spectra are well fitted by an absorbed convolution thermal-Comptonization model, almost the whole of the disk emission is up-scattered by the corona, or the boundary-layer (BL) , or the spreading-layer (SL).
The faintest burst shows a double-peaked structure and no cooling in the interval between the two sub-bursts.
For the PRE bursts, the X-ray burst shows  a significant spectral deviation/excess both at $<$ 3 keV and $>$ 10 keV from an absorbed blackbody in the PRE phase.
The residuals could be flattened by the $f_{a}$ model and the convolution  thermal Comptonization model.
For the PRE bursts, the bolometric flux of touch-down time is about twice brighter than that of the rising part of the PRE phase.

\subsection{Stalled propagation of the hot-spot during the faint burst}

The faint burst, burst \#9,  which is not a PRE burst, does not adhere to the canonical fast-rise and exponential-decay structure of most type-I X-ray bursts, instead showing a double-peaked structure.
As the bolometric flux and radius exhibit the same double-peaked profile, the temperature instead shows a plateau between the two sub-bursts.
These features are different from the double-peaked burst with about 1 s or 4 s dips with  an amplitude of 25\% or 40\% detected from 4U~1608--52 in 1984 \citep{Penninx1989} and 2017 \citep{Guver2021}, or the double-peaked burst with the radius increasing monotonically with time from GRS~1741.9--2853 \citep{Pike2021} and 4U~1636--536 \citep{Bhattacharyya2006}, or the double-peaked burst with the temperature decreasing monotonically with time during the interval of the two sub-bursts from GX~17+2 \citep{Kuulkers2002}, or the triple-peaked burst with a temperature dip during the interval of the sub-bursts \citep{Zhang2009}.

Since there is an absence of temperature dip, but with a dip in the radius between the two sub-bursts, it is natural to consider that it is due to a stalled propagation of the hot-spot, which   still burns in the stalled location, e.g., ignites on  high latitude and stalls in the equator.

It may also be a failed PRE burst, since the  sum of the two peaks has a count rate comparable to that of the PRE. In such a scenario, the entire power is released in two steps: first  by the partially burning fuel, leading to the preceding sub-burst, and then by the burning of the entire NS surface.

\subsection{Evidence of obscured NS surface during outbursts}

In theory, for the PRE bursts, there are at least two moments that the hot spot just covers the whole NS surface: the photosphere lift-up point and the touch-down  point   \citep{Shaposhnikov2003}.
Because of the fasting rise on the onset of PRE bursts, especially  most of  them with a large portion of  helium, which causes a much shorter time-scale of the rising phase;  the latter is usually used to derive the NS radius, but the former is hard to be used to derive the NS radius  due to the short rising time, i.e., the rising time is too short to   accumulate enough photons for a spectral fitting, e.g.,  PRE bursts of XTE~J1701--462 \citep{Lin2009} detected by RXTE.

Assuming the photosphere emission is isotropic,  the Eddington luminosity measured by a distant observer  is  dependent on the burning material, effective temperature and  radius of the photoshere, e.g.,  Equation 7 by \citet{Galloway}.
Based on  this equation, for the pure helium burning at the NS surface, i.e., at just the   lift-up time and the touch-down time of the photosphere, the  observed bolometric flux should be the same.
In the PRE phase, the flux observed should be higher than the above two values due to the redshift correction,
 e.g., the  observed bolometric flux with $R$=30 km is $\sim$20\% higher than that with  $R$=10 km.
In the observations of the time-resolved spectral of the PRE bursts \citep{Galloway}, the vast majority of the peak flux reached a (local) maximum close to the time of peak radius, which does follow the equation above.

The above results are based on the assumption  that the emission is isotropic, e.g., absence from   obscuration by the accretion disk or Comptonization by the corona.
However, in theory, Poynting-Robertson drag could drain the inner-accretion-disk  by   taking away  the momentum of the accretion matter hence enlarging the local accretion rate \citep{int2013,Worpel2013,Worpel2015}, which is faster than it is being refilled \citep{Stahl,Fragile2020}.
Assuming that a dynamical evolution of the disk geometry  causes this phenomenon, i.e., the lower NS hemisphere  which is obscured before the burst (the burst PRE phase) and appears from the disk after the burst-disk interaction, as shown in Figure 1 of \citet{Shaposhnikov2003} and Figure 7 of \citet{Chen2022},
the inclination angle is derived from the equation $\frac{F_{\rm rise}}{F_{\rm decay}}=(1+ {\rm cos}~i)/2$  \citep{Shaposhnikov2003,Shaposhnikov2004}, in which $F_{\rm rise}$ and  $F_{\rm decay}$  are the blackbody fluxes    detected at the rising phase and  decaying phase, respectively.

However, in this case, for the bright bursts \#3,  \#4, \#6, \#7, the ratio  $\frac{F_{\rm rise}}{F_{\rm decay}}$  are equal to or less 0.5, which is  out of the allowed range.
Since the disk could only obscure at most half of the burst emission,
 another structure between the NS surface and  our line of sight should be considered.

For the material obscuring our line of sight, a possible source is the left-over hydrogen which is not stably burned to helium.
In theory,  under a  higher accretion rate, i.e., $\sim$ 0.1--1$L_{\rm Edd}$, hydrogen  accretes faster than it can be consumed by steady burning, so that   helium ignites unstably in a H-rich environment \citep{Galloway}.
Since the persistent emission when the bursts occurred is $\sim$ 20\%$L_{\rm Edd}$, the helium bursts are expected.
The characteristic of bursts in this work does conform to the helium burning.
We know that the material in the NS surface 
is layered, e.g.,  hydrogen is on the top of   helium.
When the bursts occur, helium is ignited in the bottom, and the short time-scale of ignition and spread prevents  hydrogen ignition;  instead, the top-layered hydrogen is lifted up by the radiation pressure of the helium burning.
In this case, the lifted-up hydrogen blocks our line of  sight and causes a flux underestimation in the rising part of the PRE phase.
For the thermal emission of the  lifted-up hydrogen, its temperature is   the same as the  temperature of the NS surface, $\sim$ 0.5 keV, as our afterward estimation on the emission NS surface, which is too faint to be detected by Insight-HXMT.

We also notice that the decay phase obeys   $F \propto kT_{\rm bb}^{4}$ with the radius obtained from the touch-down time, which indicates the obscuring material is not refilled  during the decay part of the burst.
After the bursts, there should be an enhancement of soft X-ray emission--the thermal emission from the NS surface.
Assuming  the NS surface with temperature $kT_{\rm bb}$=0.5 keV and radius  $R_{\rm bb}$=6.9 km, the enhancement of the luminosity is 1/$10^{3}$ of the flux in the touch-down time, which is too faint to be detected.
Since the following burst also shows the emission shortage   during the rising PRE phase due to the obscuration,
we speculate that the obscuring material has been  rebuilt between the bursts.

Another possibility causing a different bolometric flux in the PRE phase is the different burning material, e.g., a larger portion of helium causes a higher bolometric flux of the Eddington limit.
However,   the fast-rising time scale already indicates  pure helium burning in the rising phase \citep{Lin2009}, and thus a higher helium portion is not possible.  
Yet  another possibility is that the  burst emission is anisotropic ($\xi>1$) \citep{Kuulkers2002} in the burst early phase, i.e., only part of the NS photosphere is lifted up and the rest of the photosphere is affixed to the NS surface.
To a distant observer,
thus can create the illusion of all of the photosphere were lifted up.
However, it is hard to explain that the flux of the burst when the photospheric radius reaches its peak is still lower than that of the touch-down time.
Especially, in theory, a smaller red-shift correction when the photospheric radius reaches its peak would cause a higher flux than that of the touch-down time.

Since the effective area of the X-ray telescopes in orbit is not big enough to detect  the rebuilding  of the  obscuring material or the process of  obscuring the NS surface, a  larger  detection area   may be satisfied by the next generation of Chinese mission of the  so-called eXTP (enhanced X-ray Timing and Polarimetry mission) \citep{Zhang2019} or  by stacking the lightcurves/spectra of the interval between bursts under the circumstance of enough bursts and   relatively smooth persistent emission.

\subsection{Comptonization of the burst emission by the  BL\&SL}

Regarding that the  temperature and optical depth deviate from the corona's  canonical values,  we   prefer another corona pattern, the  so-called warm layer \citep{Zhang2000} with temperature $\sim$2--3 keV and  optical depth $\sim$5--10, which is produced by  the magnetic reconnection.
The Comptonization of the burst emission by the BL/SL, by adopting the convolution  model (with an input seed photon spectrum of blackbody)  with the parameters derived in the persistent emission fitting, was first adopted in the burst detected from 4U~1608--52 \citep{Chen2022}.
Similar to our previous work \citep{Chen2022}, the scenario is also  applied to the five bursts from 4U~1730--22, resulting in an  equally good fit compared with the $f_{a}$ model during the bright/PRE phase.
Under this scenario,  the radius of the photosphere is underestimated with the canonical blackbody model or the  $f_{a}$ model.

\section{Summary and Conclusion}

In summary, from Insight-HXMT observations on 4U~1730--22, we present here  one non-PRE burst with the double-peaked profile and  no cooling  between the two sub-bursts, and nine PRE bursts with the flux shortage during the rising  phase, which can be attributed to  a stalled burning front at the equator, and  occlusion by the material in our line of sight or an anisotropic  emission  in the burst early phase, respectively.


\acknowledgements
This work made use of the data and software from the Insight-HXMT
mission, a project funded by China National Space Administration
(CNSA) and the Chinese Academy of Sciences (CAS).
This research has  made use of data and software provided by of data obtained from the High Energy Astrophysics Science
Archive Research Center (HEASARC), provided by NASA’s
Goddard Space Flight Center.
This work is supported by the National Key R\&D Program of China (2021YFA0718500) and the National Natural Science Foundation of China under grants 11733009, U1838201, U1838202, U1938101, U2038101, 12130342, U1938107.

\bibliographystyle{plainnat}


\begin{table}[ptbptbptb]\small
\begin{center}
\caption{The bursts obsid and peak time of 4U~1730--22  detected by Insight-HXMT in 2021 \& 2022 outbursts}
\label{tb_burst}
\begin{tabular}{cccccccccccccccccc}
\\\hline
  No & obsid & Burst peak time   & $F_{\rm peak}$ & $E_{\rm b}$ &PRE \\\hline
   &    	&  MJD  & $10^{-8}~{\rm erg}~{\rm cm}^{2}~{\rm s}^{-1}$ & $10^{-8}~{\rm erg}~{\rm cm}^{2}$  &  \\\hline
1&P041401100410-20210709-02-01$^{*}$	& 59404.30775  & 3.4$\pm$0.3 & 26.9$\pm$1.0&Y  	\\\hline
2&P041401100801-20210716-01-01$^{*}$ & 59411.72027 & 4.6$\pm$0.4 & 56.6$\pm$1.5&Y  	\\\hline
3&P051400200102-20220430-01-01 & 59699.26105   & 4.2$\pm$0.2   & 30.8$\pm$0.5&Y  	\\\hline
4&P051400200402-20220503-01-01 &59702.29225& 3.0$\pm$0.2 & 25.3$\pm$0.4&Y \\\hline
5&P051400200601-20220505-01-01 &59704.34837& 2.9$\pm$0.2  &  32.7$\pm$0.5&Y \\\hline
6&P051400200701-20220506-01-01 &59705.15499&  4.2$\pm$0.2  &  33.5$\pm$0.5&Y  \\\hline
7&P051400200801-20220507-01-01 &59706.25535&  4.9$\pm$0.3   &  42.2$\pm$0.8 &Y  \\\hline
8&P051400200902-20220508-01-01$^{*}$ &59707.34785&  3.7$\pm$0.4   &  27.0$\pm$0.8&Y  \\\hline
9&P051400201003-20220509-01-01$^{d}$ &59708.45082&  1.7$\pm$0.2   &  12.7$\pm$0.4&N \\\hline
10&P051400201102-20220510-01-01$^{*}$ &59709.33134&  3.9$\pm$0.4   &  25.5$\pm$0.8&Y \\\hline
\end{tabular}
\end{center}
\begin{list}{}{}
\item[$^{*}$]{The bursts only detected by LE}
\item[$^{d}$]{The burst with double-peak profile}
\end{list}
\end{table}

\begin{table}[ptbptbptb]\small
\begin{center}
\caption{The NICER obsid on the same day when Insight-HXMT detected burst \#3 }
 \label{tb_nicer_obsid}
\begin{tabular}{cccccccccccc}
\\\hline
obsid &  Start Time & 	GTI (s)    \\\hline
4639010134  &  59700.04657 (2022-05-01T01:02:20)  &453\\\hline
\hline
\end{tabular}
\end{center}
\end{table}

\begin{table}
\centering
\caption{The results of the spectral fit of the LE, ME and NICER spectra in the 0.4--20 keV range   with  cons*tbabs*thcomp*diskbb}
\label{tb_persist_fit}
\vskip -0.4cm
\begin{tabular}{ccccccc}
\\\hline
$N_{\rm H}$  & $\tau$ & $kT_{\rm e}$  & $f_{\rm sc}$ & $kT_{\rm in}$
 & $N_{\rm diskbb}$& $\chi_\nu^2$ \\
$10^{22}~{\rm cm}^{-2}$ &  & keV & &keV &  &   \\
\hline
$0.53_{-0.01}^{+0.01}$ & $7.8^{+1.0}_{-0.9}$ & $3.61^{+0.42}_{-0.57}$ & $0.929_{-0.12}^{}$  & $0.68_{-0.34}^{+0.39}$ & $370.3_{-62.8}^{+74.0}$ & 146.1/160 \\ 

\hline
\end{tabular}
\end{table}


\clearpage


\clearpage

\clearpage

\clearpage

\end{document}